%% file: sample-sigconf.tex
\documentclass[sigconf]{acmart}

\usepackage{hyperref}
\usepackage{url}

\usepackage{color}
\usepackage{pifont}

\usepackage{graphicx} 
\usepackage{makecell}
\usepackage{amsmath}

\usepackage{xcolor}

\usepackage{multirow}
\usepackage{subfigure}

\usepackage{soul}

\usepackage[ruled]{algorithm2e}

\newtheorem{prop}{Proposition}
\newtheorem{assume}{Assumption}
\newtheorem{lemma}{Lemma}
\newtheorem{theorem}{Theorem}
\newtheorem{remark}{Remark}

\newcommand{\E}{\mathbb{E}}
\newcommand{\Var}{\mathrm{Var}}

\newcommand{\normlone}{L^1}
\newcommand{\normltwo}{L^2}

\AtBeginDocument{%
  \providecommand\BibTeX{{%
    \normalfont B\kern-0.5em{\scshape i\kern-0.25em b}\kern-0.8em\TeX}}}

\setcopyright{acmcopyright}
\copyrightyear{2018}
\acmYear{2018}
\acmDOI{XXXXXXX.XXXXXXX}

\acmConference[KDD '23]{29th ACM SIGKDD Conference on Knowledge Discovery and Data Mining}{August 6-10, 2023}{Long Beach, CA, USA}
\acmPrice{15.00}
\acmISBN{978-1-4503-XXXX-X/18/06}

\begin{document}

\title{Robust Quantity-Aware Aggregation for Federated Learning}
\renewcommand{\shorttitle}{Robust Quantity-Aware Aggregation for Federated Learning}
\author{Jingwei Yi$^1$, Fangzhao Wu$^2$, Huishuai Zhang$^2$, Bin Zhu$^2$, Guangzhong Sun$^1$, Xing Xie$^2$}
\renewcommand{\authors}{Jingwei Yi, Fangzhao Wu, Huishuai Zhang, Bin Zhu, Guangzhong Sun, Xing Xie}
\affiliation{%
  \institution{$^1$University of Science and Technology of China, Hefei 230027, China  \\
  $^2$Microsoft Research Asia, Beijing 100080, China}
  \country{}
} 
\email{yjw1029@mail.ustc.edu.cn, gzsun@ustc.edu.cn,{fangzwu,huzhang,binzhu,xingx}@microsoft.com}

\renewcommand{\shortauthors}{Yi, et al.}

\input{src/abstract}






\input{src/abstract}
\maketitle

\input{src/intro}
\input{src/related}

\input{src/problem}
\input{src/attack}
\input{src/existing}
\input{src/method}

\input{src/theory}
\input{src/exp}

\input{src/conclusion}

\bibliographystyle{ACM-Reference-Format}
\bibliography{main}

\clearpage
\appendix

\input{src/supp.tex}

\end{document}

%% file: src/abstract.tex
\begin{abstract}

Federated Learning (FL) enables multiple clients to collaborate on training machine learning models without sharing their data, making it an important privacy-preserving framework. 
However, FL is susceptible to security and robustness challenges, such as malicious clients potentially corrupting model updates and amplifying their impact through excessive quantities.
Existing defenses for FL, while all handling malicious model updates, either treat all quantities as benign or ignore/truncate quantities from all clients.
The former is vulnerable to quantity-enhanced attacks, while the latter results in sub-optimal performance due to the varying sizes of local data across clients.
In this paper, we introduce a robust quantity-aware aggregation algorithm for FL called FedRA. FedRA considers local data quantities during aggregation and defends against quantity-enhanced attacks.
It consists of two crucial components, i.e., a quantity-robust scorer and a malicious client number estimator. 
More specifically, the quantity-robust scorer calculates the sum of the distance between one update and the rest of the updates as the quantity-robust score, taking into account their quantities as factors.
The malicious client number estimator uses these scores to predict the number of suspicious clients to exclude in each round, adapting to the dynamic number of malicious clients participating in FL in each round.
Experiments on three public datasets demonstrate FedRA's effectiveness in defending against quantity-enhanced attacks in FL. The code for FedRA can be found at \url{https://anonymous.4open.science/r/FedRA-4C1E}.


\end{abstract}

%% file: src/intro.tex
\section{Introduction}
Federated learning (FL) is an important technology to train models while protecting the privacy of training data.
It has been extensively studied for various application scenarios, such as medical health~\citep{rieke2020future,sheller2020federated,xu2021federated,nguyen2022federated,kumar2021federated} and keyboard next-word prediction~\citep{yang2018applied,hard2018federated,chen2019federated,li2020review,ramaswamy2019federated}.
One of the classic FL algorithms is FedAvg~\citep{mcmahan2017communication}.
In FedAvg, the server iteratively averages clients' updates with some weights determined by the \emph{quantity} of each client, which means throughout the paper the number of the training data at that client, to update the global model.

The linear aggregation used in FedAvg has been shown to be vulnerable to poisoning attacks~\citep{NEURIPS2019_ec1c5914,10.5555/3489212.3489304,DBLP:conf/nips/WangSRVASLP20,liu2017trojaning,bagdasaryan2020backdoor,bhagoji2019analyzing,xie2021crfl}. These attacks primarily aim to either deteriorate the performance of the global model or insert a backdoor into the global model.
However, malicious clients can also submit large quantities to obtain unfairly high weights in the model aggregation, leading to a heightened impact of their malicious updates on the global model.
We refer to this type of attack as \textbf{quantity-enhanced attacks}.


Several methods have been proposed to defend against poisoning attacks for federated learning~\citep{blanchard2017machine,yin2018byzantine,pmlr-v80-mhamdi18a,sun2019can,pillutla2019robust,portnoy2020towards}.
These defenses can be categorized into three groups: quantity-ignorant, quantity-aware, and quantity-robust defenses.
\textbf{Quantity-ignorant} defenses aggregate updates without considering quantities~\citep{blanchard2017machine, yin2018byzantine, pmlr-v80-mhamdi18a}.
These methods are robust to quantity-enhanced attacks.
However, since aggregating updates with quantities benefits model performance~\citep{zaheer2018adaptive,reddi2021adaptive}, applying these defenses may lead to performance degradation.
\textbf{Quantity-aware} defenses aggregate updates with quantities but by default treat quantities submitted by clients as benign~\citep{sun2019can, pillutla2019robust}.
These defenses outperform quantity-ignorant defenses when there are no attacks, but their performance deteriorates significantly when faced with quantity-enhanced attacks.
\textbf{Quantity-robust} defenses aggregate updates with quantities and, unlike quantity-aware defenses, are robust to quantity-enhanced attacks.
~\citet{portnoy2020towards} propose to truncate quantities with a dynamic threshold and apply quantity-aware Trimean~\citep{yin2018byzantine} to aggregate updates.
However, it handles updates and quantities separately, which could result in the truncation of the quantities of benign clients, ultimately leading to sub-optimal performance.

Meanwhile, some existing defenses~\citep{blanchard2017machine, yin2018byzantine, pmlr-v80-mhamdi18a} need a parameter that represents the upper bound of the number of malicious clients to be filtered out in each round.
However, in cross-device federated learning, due to the large number of clients, only a subset of clients is selected for participation in each round.
As a result, 
the number of malicious clients in each round changes dynamically.
Over-estimating the number of malicious clients  leads to some benign clients being filtered out, while underestimating the number of malicious clients may result in some malicious updates being included in the model aggregation.

In this paper, we propose \emph{FedRA}, a quantity-robust defense for federated learning. 
It filters out malicious clients by taking both quantities and updates into consideration.
More specifically, FedRA consists of two components, i.e., quantity-robust scorer and malicious client number estimator.
The quantity-robust scorer is based on the observation that the expectation of distance between benign updates with larger quantities should be smaller.
Therefore, to compute the quantity-robust score, we calculate the sum of distances between one update and the rest of the updates, taking into account the quantities as coefficients.
The coefficient is smaller when the pair of quantities are larger.
The malicious client number estimator uses the quantity-robust scores as input and predicts the number of malicious clients to exclude in each round by maximizing the log-likelihood.
Finally, the corresponding number of updates with the largest quantity-robust scores are filtered out.
The rest updates are aggregated with weights proportional to their quantities to update the global model.

The main contributions of this paper are as follows:
\begin{itemize}
    \item We propose a robust quantity-aware aggregation method for federated learning to aggregate updates with quantities while defending against quantity-enhanced attacks.
    \item We theoretically prove FedRA is quantity-robust by proving its aggregation error is irrelevant to malicious quantities.
    \item We conduct experiments on three public datasets to validate the effectiveness of FedRA.
\end{itemize}

%% file: src/related.tex
\section{Related Works}
\subsection{Federated Learning}
Federated learning~\citep{mcmahan2017communication} enables multiple clients collaboratively train models without sharing their local datasets.
There are three steps in each round of federated learning.
First, a central server randomly samples a group of clients and distributes the global model to them.
Second, the selected clients train the model with their local datasets and upload their model updates to the central server.
Finally, the central server aggregates the received updates to update the global model.
In FedAvg~\citep{mcmahan2017communication}, updates are weight-averaged according to the quantity of each client's training samples.
~\citet{reddi2021adaptive} have proposed FedAdam, where the server updates the global model with Adam optimizer.
The above steps are performed iteratively until the global model converges.

\subsection{Federated Poisoning Attacks}
However, classical federated learning is vulnerable to poisoning attacks, e.g., untargeted attacks~\citep{NEURIPS2019_ec1c5914,10.5555/3489212.3489304} and backdoor attacks~\citep{liu2017trojaning,bagdasaryan2020backdoor,DBLP:conf/nips/WangSRVASLP20,Xie2020DBA:,bhagoji2019analyzing}.
In this paper, we focus on defending against untargeted attacks, which aim to degrade the performance of the global model on arbitrary input samples.
Label flip~\cite{10.5555/3489212.3489304} is a data-poisoning-based untargeted attack, where the malicious clients manipulate the labels of their local training dataset to generate malicious updates.
Some advanced attacks assume the malicious clients have knowledge of benign datasets and can collaborate to generate malicious updates.
LIE~\cite{NEURIPS2019_ec1c5914} adds small noises to the average of benign updates to circumvent defenses.
Optimize~\cite{10.5555/3489212.3489304} models the attack as an optimization problem and adds noises in the opposite direction of benign updates.
These attacks focus on generating malicious updates, but overlook the attacker can submit large quantities to obtain unfairly high weight in model aggregation.

\subsection{Robust Aggregations}
To defend against untargeted attacks in federated learning, several robust aggregation methods have been proposed.
~\citet{yin2018byzantine} propose Median and Trimean that apply coordinate-wise median and trimmed-mean, respectively, to filter out malicious updates.
~\citet{blanchard2017machine} propose Krum and mKrum that compute square-distance-based scores to select and average the updates closest to a subset of neighboring updates. 
Bulyan~\citep{pmlr-v80-mhamdi18a} is a combination of mKrum and Trimean: it first selects several updates through mKrum and then aggregates them with Trimean.
We note that the above defense methods do not consider quantities of clients' training samples, which we categorize as \emph{quantity-ignorant} methods, and the convergence speeds and model performance of these methods are compromised~\citep{zaheer2018adaptive,reddi2021adaptive}, especially for quantity-imbalanced scenarios, such as long-tailed data distributions that are common in real-world scenarios~\citep{Zhang_2017_ICCV,Zhong_2019_CVPR,10.1145/3123266.3123316}.

\citet{sun2019can} propose Norm-bound that clips the $\normltwo$ norm of received updates to a predefined threshold.
~\citet{pillutla2019robust} propose RFA that computes weights for each update by running an approximation algorithm to minimize the quantity-aware geometric median of updates.
These two methods are categorized into \emph{quantity-aware} defenses. They take quantities into consideration when aggregating updates but by default treat all received quantities as benign. 
\citet{portnoy2020towards} point out that received quantities may be malicious and can be exploited to increase the impact of malicious updates.
They further propose a Truncate method that truncates received quantities within a dynamic threshold in each round, which guarantees any 10\% clients do not have more than 50\% samples.
The Truncate method is categorized into \emph{quantity-robust} method.
However, quantities of benign clients with a large number of training samples may also be truncated, resulting in degraded performance.
Meanwhile, they handle the malicious update filtering and the quantity truncation separately, which is sub-optimal.

%% file: src/problem.tex
\begin{figure*}[!t] 
    \centering
    \subfigure[MNIST in IID setting]{\includegraphics[width=0.25\textwidth]{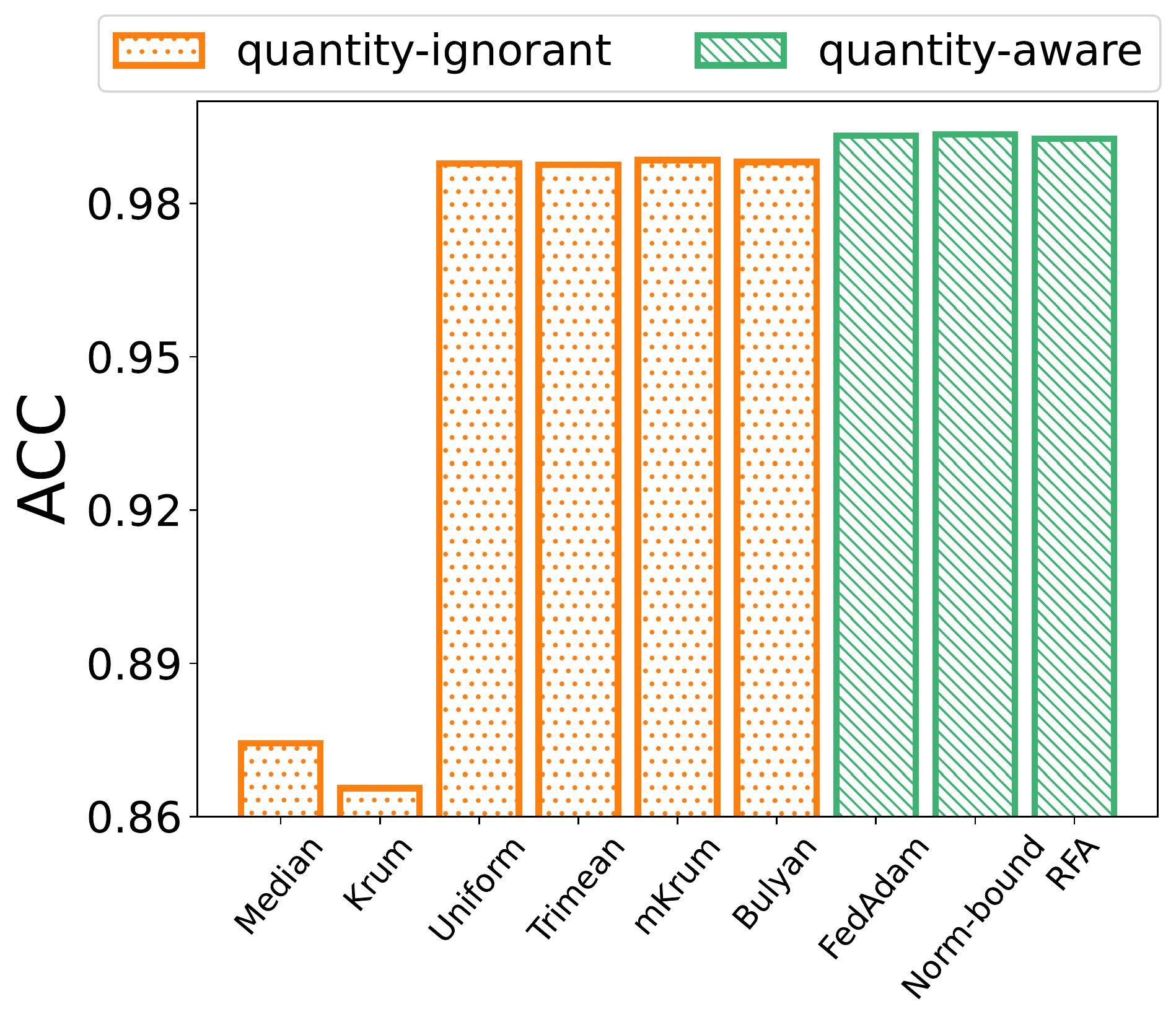}}
    \subfigure[CIFAR10 in IID setting]{\includegraphics[width=0.25\textwidth]{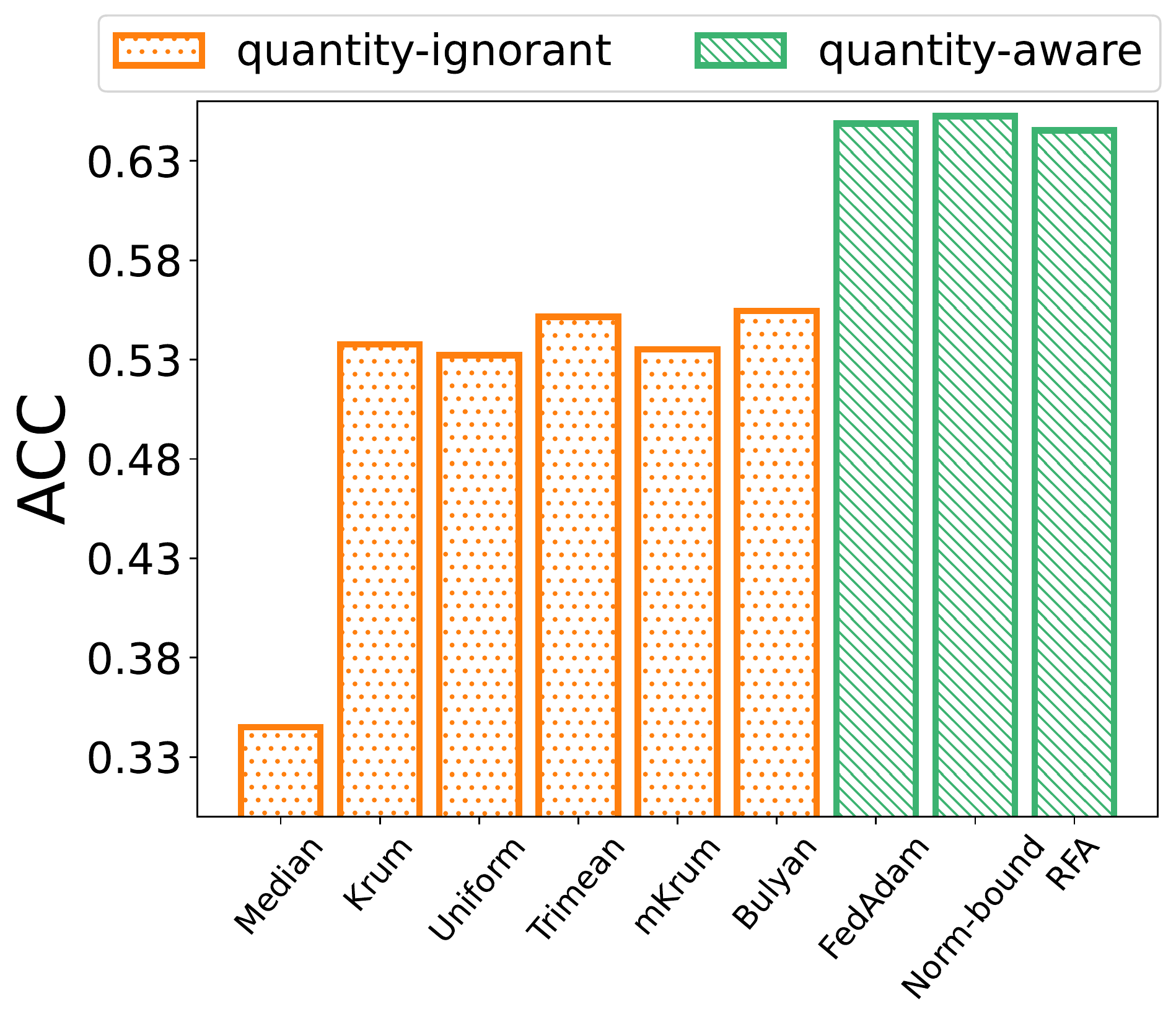}}
    \subfigure[MIND in IID setting]{\includegraphics[width=0.25\textwidth]{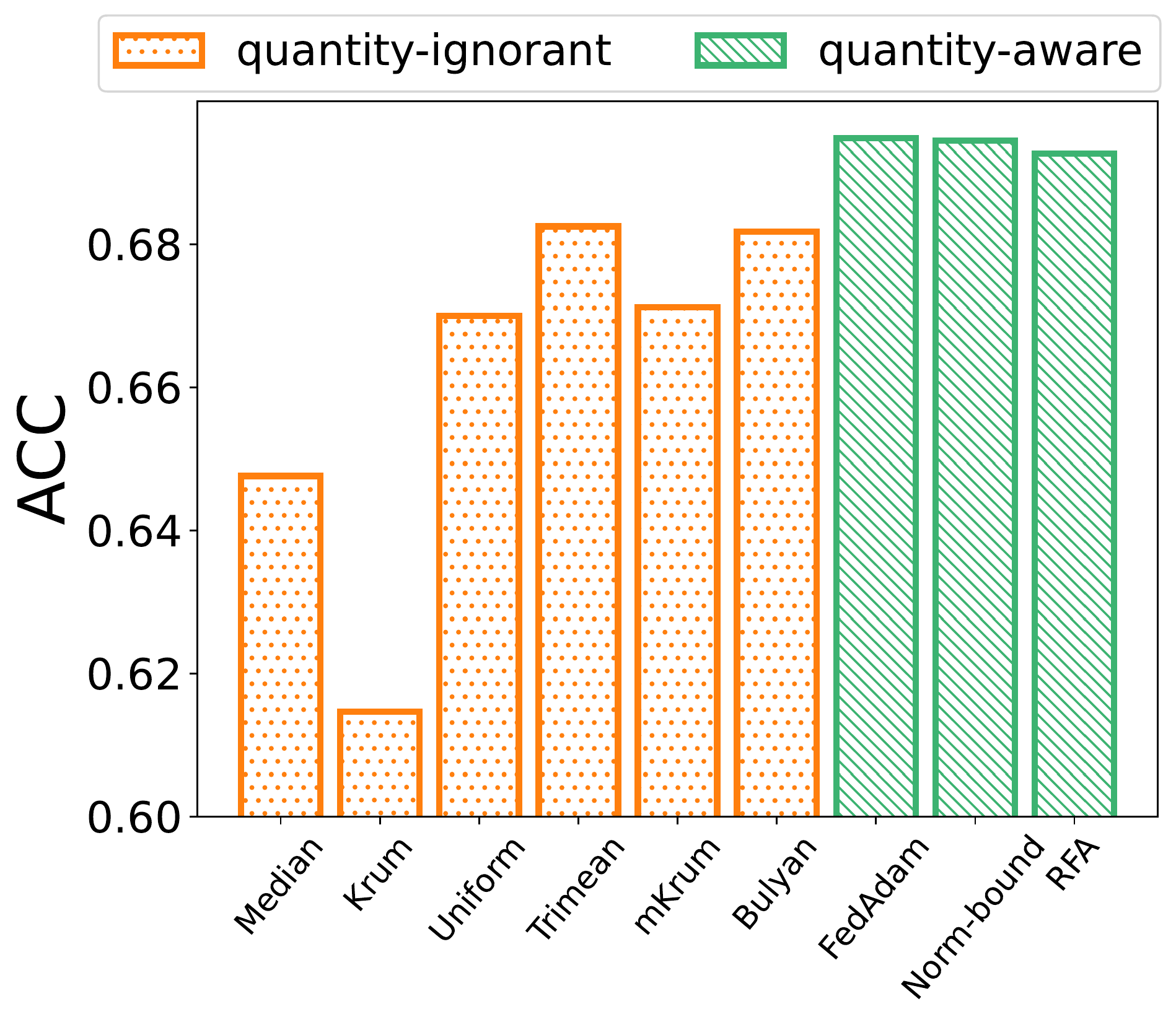}}
    \subfigure[MNIST in non-IID setting]{\includegraphics[width=0.25\textwidth]{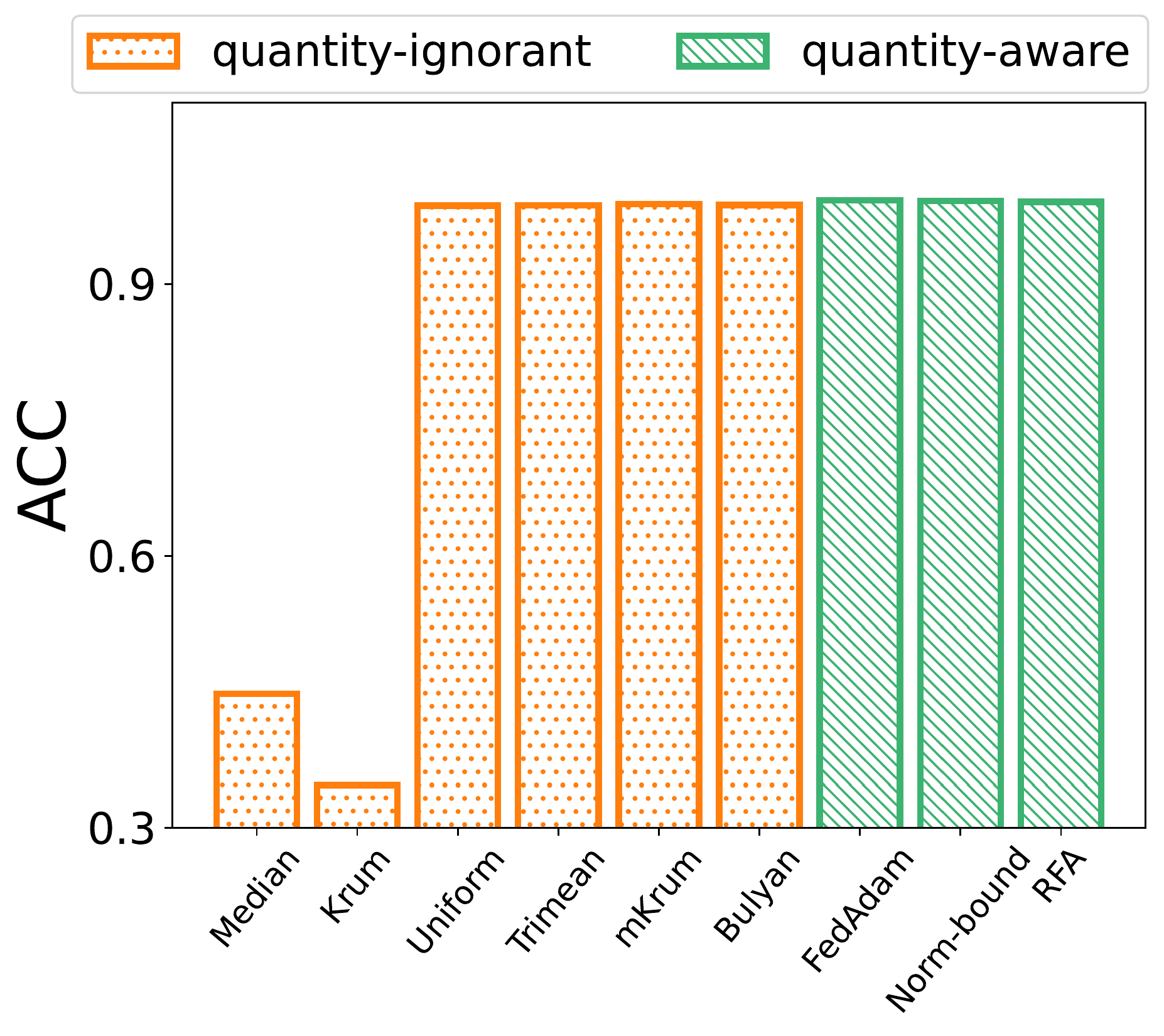}}
    \subfigure[CIFAR10 in non-IID setting]{\includegraphics[width=0.25\textwidth]{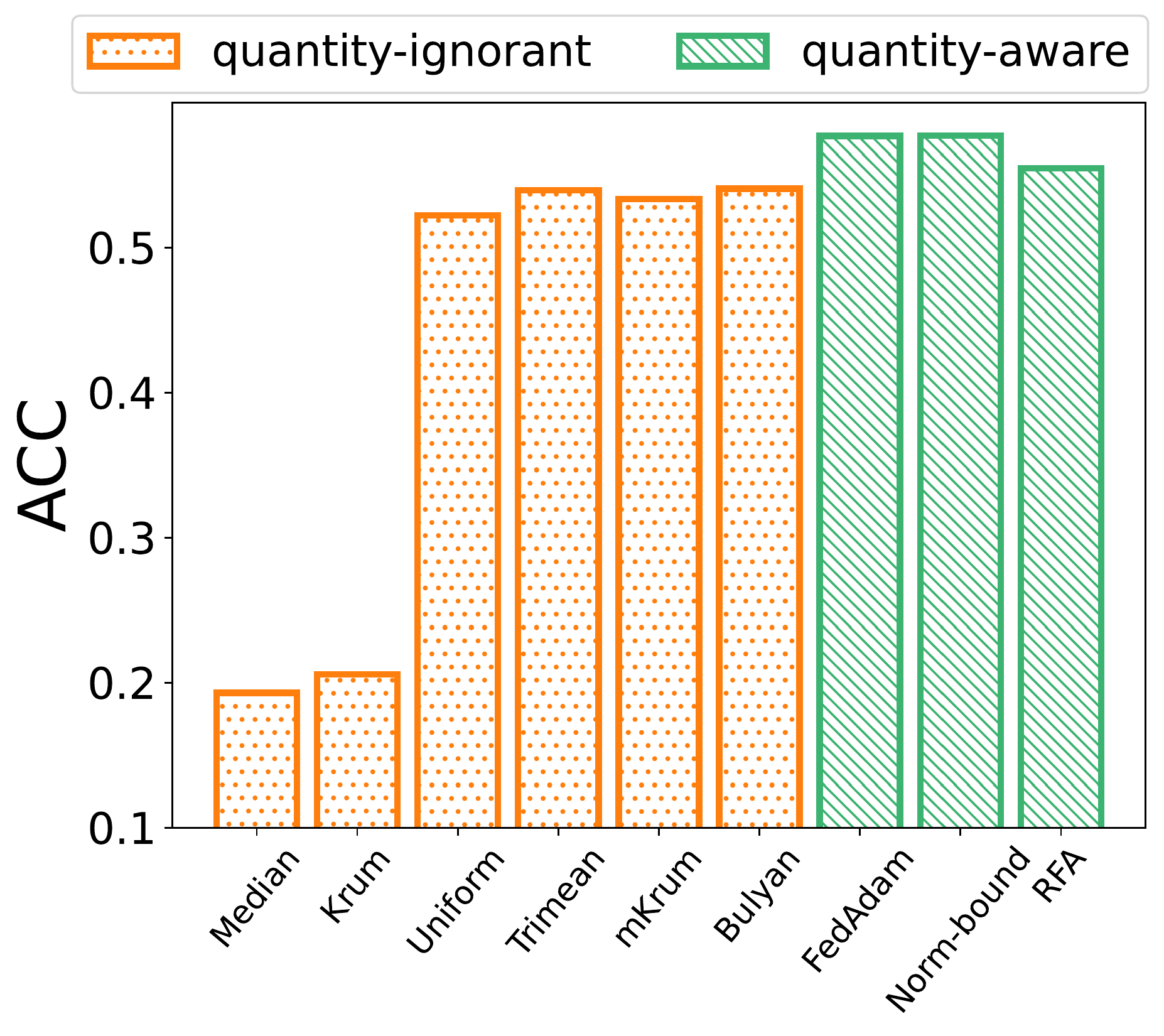}}
    \subfigure[MIND in non-IID setting]{\includegraphics[width=0.25\textwidth]{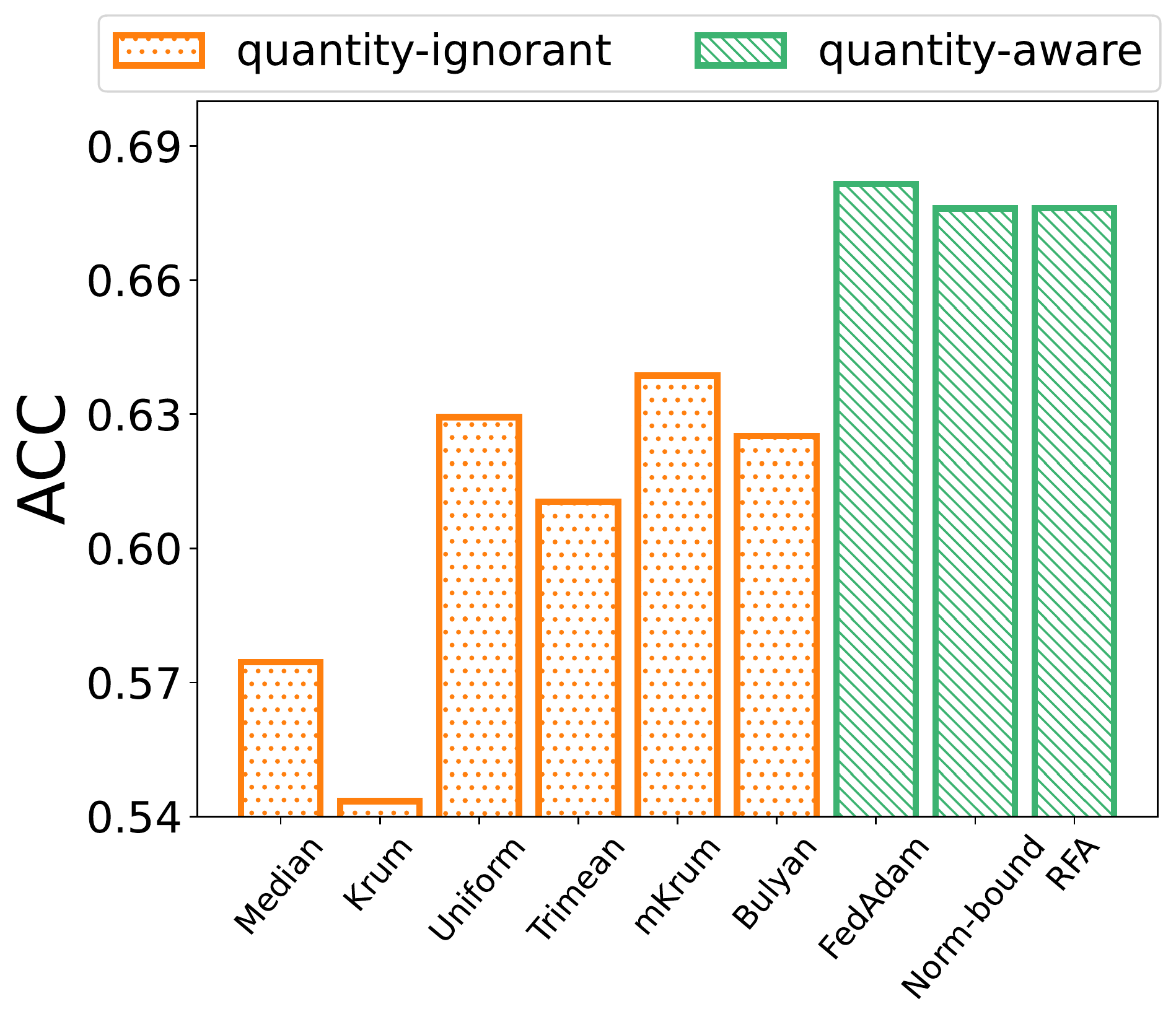}}
    \caption{Quantity-aware methods outperform quantity-ignorant methods on three datasets in both IID and non-IID settings.}
    \label{fig:no-attack}
\end{figure*}
\begin{figure}[!t] 
  \centering
  \subfigure[MNIST in fixed-ratio IID setting]{\includegraphics[width=0.23\textwidth]{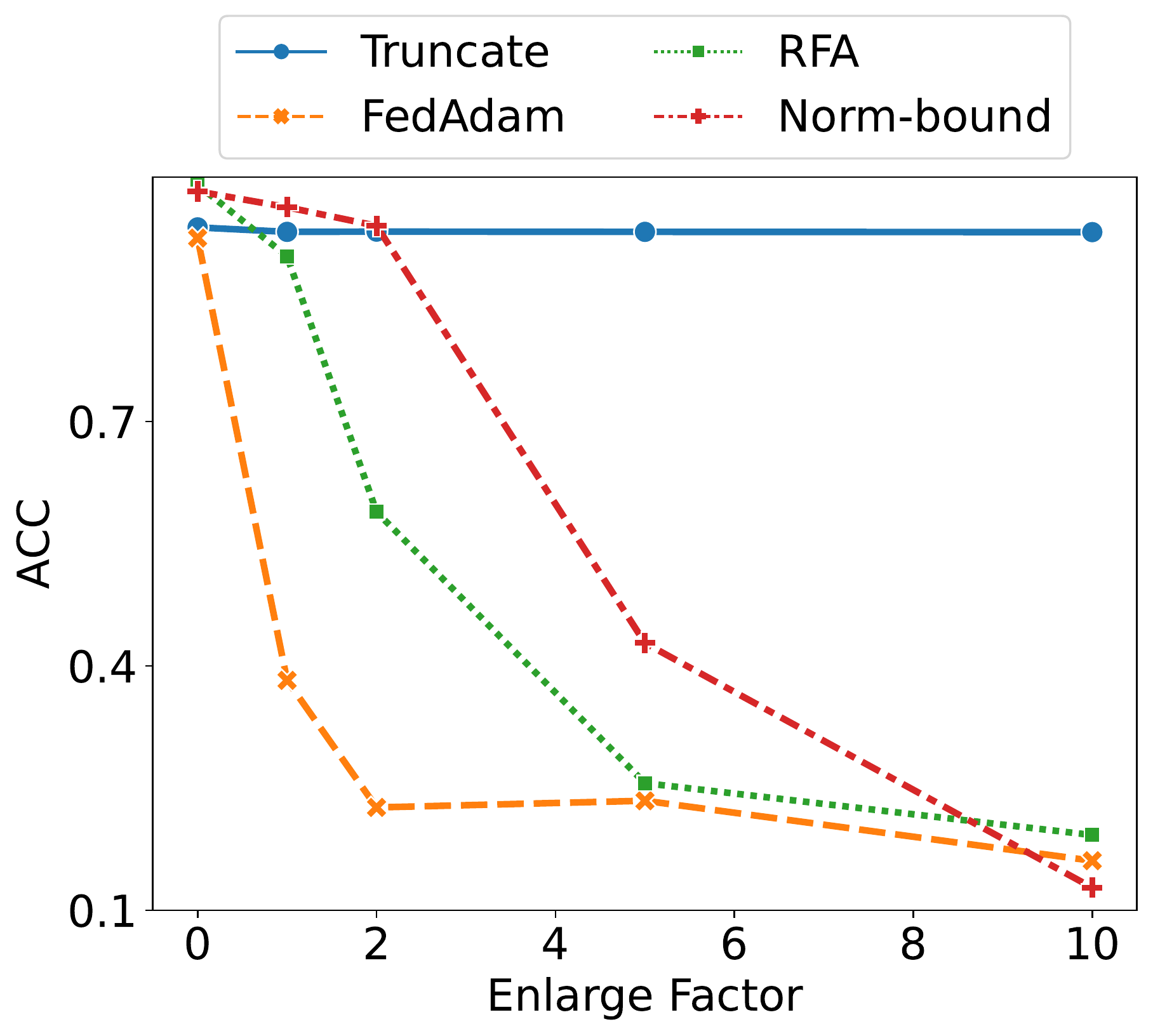}}
  \subfigure[MNIST in fixed-ratio non-IID setting]{\includegraphics[width=0.23\textwidth]{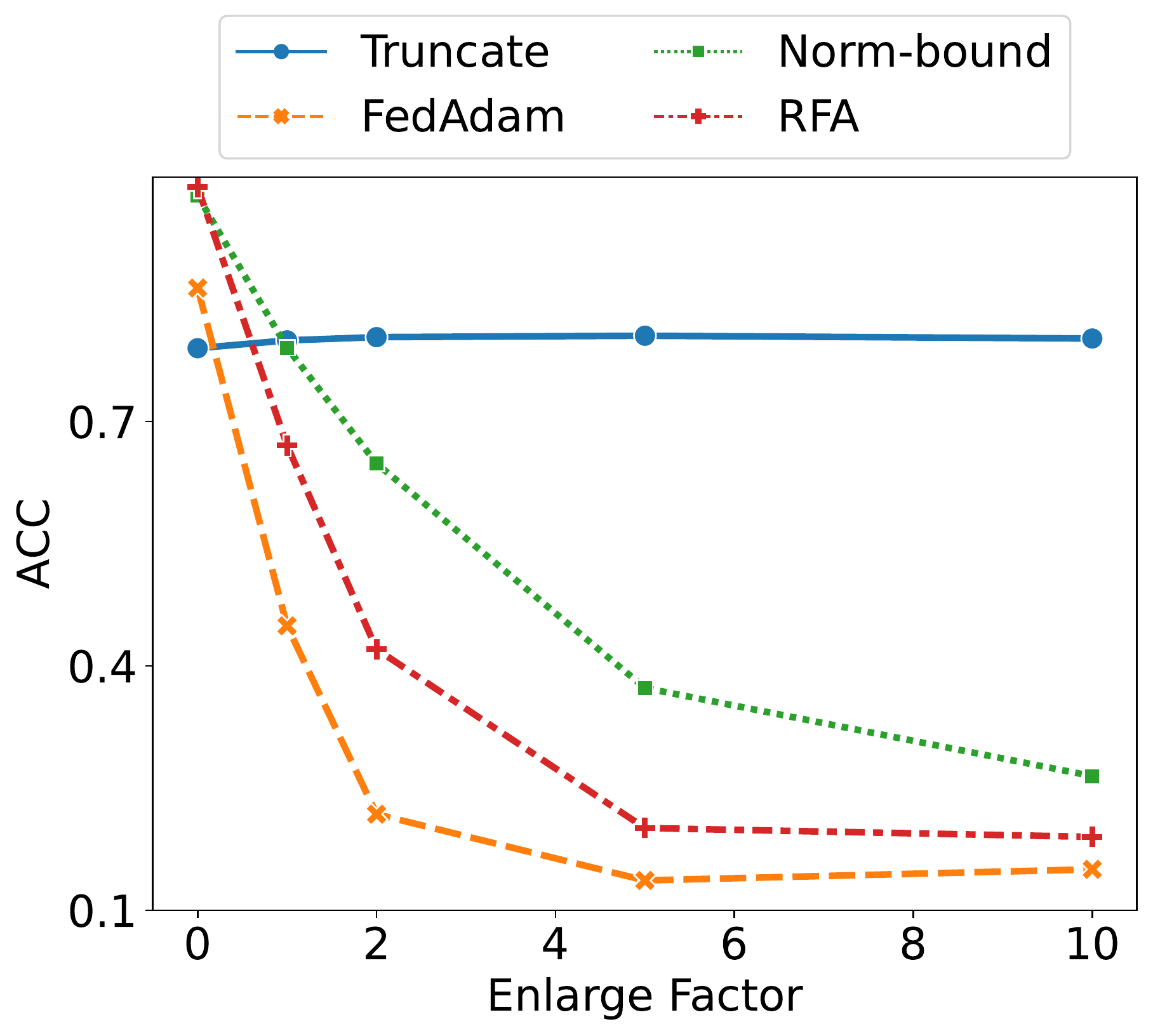}}
  \caption{Quantity-aware methods are not robust to LIE-based quantity-enhanced attacks.}
  \label{fig:lie-mnist-quantity-aware}
\end{figure}


\section{Problem Definition}
\label{sec:prob-define}
Suppose that training samples are sampled from a distribution $\mathcal{D}$ in sample space $\mathcal{Z}$.
Let $f(\textbf{w};z)$ denote the loss function of model parameter $\textbf{w}\in \mathcal{W}$ at data point $z$, and $F(\textbf{w})=\E_{z\sim\mathcal{D}}[f(\textbf{w};z)]$ is the corresponding population loss function.
The goal is to  minimize the population loss by training the model parameter, i.e., $\textbf{w}^* = \arg\min_{\textbf{w}\in \mathcal{W}}F(\textbf{w})$.

Assume that there are $N$ clients in total and $M$ of them are malicious.
The $i$-th client has a local dataset $D_i$.
The empirical loss of the $i$-th client is $F_i(\textbf{w}) = \frac{1}{|D_i|}\sum_{z\in D_i}f(\textbf{w};z)$.
In the $t$-th round, the central server randomly samples $n$ clients and distributes the global model $\textbf{w}_t$ to them.
Following existing works~\cite{yin2018byzantine, blanchard2017machine, pmlr-v80-mhamdi18a}, we define a benign client will submit update $\textbf{g}_t^i = \nabla F_i(\textbf{w}_t)$ and quantity $q^i_t = |D_i|$, while a malicious client can submit an arbitrary update and an arbitrary quantity to the server.
After receiving the updates and quantities from the sampled $n$ clients, the server computes the global update with a certain aggregation rule $\mathcal{A}$: $\textbf{g}_{t+1} =  \mathcal{A}(\textbf{g}_t^1, ... , \textbf{g}_t^n, q^1_t, ... , q^n_t)$.

Some existing robust aggregation algorithms, e.g., mKrum~\citep{blanchard2017machine}, Bulyan~\citep{pmlr-v80-mhamdi18a}, and Trimean~\citep{yin2018byzantine}, require estimation of the upper bound of the number of malicious clients, denoted by $\widetilde{m}$, in each round using a fixed parameter for all rounds.
However, in cross-device federated learning, due to the large number of clients, only a subset of clients is selected for participation in each round.
The number of malicious clients $m$ in each round follows a hypergeometric distribution and is thus hard to be estimated by a fixed parameter.
Over-estimating the number of malicious clients leads to some benign clients being filtered out, while underestimating the number of malicious clients may result in some malicious updates being included in the model aggregation.
In our work, we consider two settings: fixed-ratio setting and dynamic-ratio setting. 
In the fixed-ratio setting, the number of malicious clients $m$ in each round is fixed, i.e., $m = \lceil\frac{nM}{N}\rceil$.
Since $m$ is not a random variable, estimating $m$ with a fixed parameter is feasible.
The fixed-ratio setting, while not representative of real-world cross-device federated learning, provides a baseline for the upper limit of defense performance in filtering out malicious updates.
In contrast, the dynamic-ratio setting involves a fixed total number of malicious clients, $M$, but the exact number of malicious clients per round is unknown, which better reflects real-world federated learning situations.

%% file: src/attack.tex
\begin{figure*}
  \centering  \includegraphics[width=0.80\textwidth]{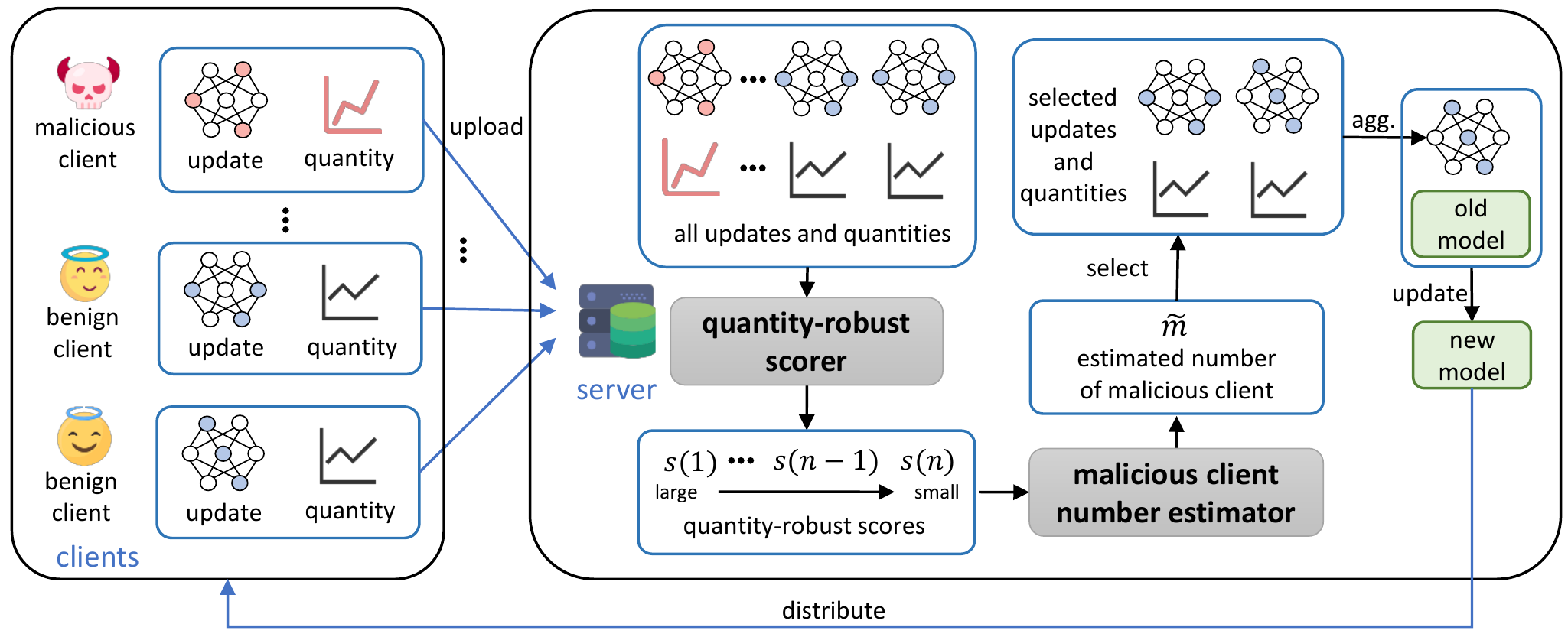}
  \caption{The framework of our FedRA.}
  \label{fig:framework}
\end{figure*}
\section{Threat Model}
Following existing works~\cite{10.5555/3489212.3489304}, in this section, we introduce the objective, knowledge, and capability of attackers and the details of quantity-enhanced attacks.

\noindent \textbf{Attacker's Objective.} The attacker's objective is to carry out untargeted attacks on the federated learning system, which aims to degrade the overall performance of the global model on arbitrary input samples. 

\noindent \textbf{Attacker's Knowledge.}
Since the server distributes the global model to clients in each round, the attacker knows the model structure, the local model parameters, and the local training code.
Due to the privacy-protecting nature of federal learning, the attacker cannot access the datasets of other clients.
Thus, the attacker has partial knowledge of the data distribution.
Since the server might not release its aggregation code, the attacker does not know the aggregation rule applied by the server.

\noindent \textbf{Attacker's Capability.}
Building on prior attack methods~\cite{NEURIPS2019_ec1c5914,10.5555/3489212.3489304}, we consider the scenario where an attacker controls $M$ malicious clients.
These clients can coordinate with each other and use benign data stored on their devices to generate malicious updates and quantities.
If any of the malicious clients are selected by the server, they can upload the malicious updates and quantities to the server, degrading the performance of the global model.

\noindent \textbf{Quantity-enhanced Attack.}
In addition to generating malicious updates via existing untargeted attacks, the attacker also submits large quantities to obtain unfairly high weight in model aggregation, resulting in an amplified impact of malicious updates on the global model.
The malicious clients first collaborate to compute the mean and standard deviation of their quantities, which are denoted as $\mu_q^m$ and $\sigma_q^m$, respectively.
The final submitted quantity is defined as $\mu_q^m + \alpha_q \sigma_q^m$, where $\alpha_q$ is the quantity-enlarging factor.
$\alpha_q$ should be as large as possible but not detectable by the server

%% file: src/existing.tex
\section{Existing Defenses Are Sub-optimal in Quantity-imbalanced Scenarios}
In this section, we prove that both quantity-ignorant and quantity-aware defenses are sub-optimal in quantity-imbalanced scenarios.
To simulate quantity-imbalanced settings, we make the quantities of the benign client datasets obey log-normal distributions, which is a typical common long-tailed distribution in real-world scenarios~\cite{li2019both,park2008long,chen2020esam,kim2019sequential,long2022retrieval}.
The detailed experimental settings are in Section~\ref{sec:exp}.

\subsection{Quantity-ignorant Methods}
In this subsection, we compare the performance of quantity-ignorant methods (Krum, mKrum, Median, TriMean, Bulyan, and Uniform) with that of quantity-aware methods (FedAdam, Norm-bounding, and RFA) in both IID and non-IID non-attack settings.
The experimental results on MNIST, CIFAR10, and MIND are shown in Figure~\ref{fig:no-attack}.
We can observe that the performance of quantity-ignorant methods is consistently lower than that of quantity-aware methods.
This is because the expectation of the local model error decreases as the sample size of the training dataset increases~\cite{ruder2016overview}.
Assigning higher weight to the local models trained with larger datasets reduced the aggregation error of the global model.

\subsection{Quantity-aware Methods}
In this subsection, we analyze the performance of quantity-aware methods (FedAdam, Norm-bounding, and RFA) in IID fixed-ratio settings when exposed to quantity-enhanced attacks.
We set the quantity-enlarging factor $\alpha_q \in \{0, 1, 2, 5, 10\}$.
Due to the space limitation, we show the results of LIE-based quantity-enhanced attacks on MNIST datasets in Figure~\ref{fig:lie-mnist-quantity-aware}.
We can observe that the performance of quantity-aware methods degrades as the quantity-enlarging factor $\alpha_q$ increases.
This is because these methods by default treat the quantities submitted by clients as benign, which is vulnerable to quantity-enhanced attacks since the large quantities can amplify the impact of malicious updates.

We also show the performance of Truncate in Figure~\ref{fig:lie-mnist-quantity-aware}.
We can observe that its performance is stable but sub-optimal under the quantity-enhanced attack.
This is because although Truncate limits the upper bound of malicious quantities, it also clips the quantities of benign clients.
Additionally, the separate handling of malicious quantities and updates restricts Truncate's ability to effectively filter malicious updates.

%% file: src/method.tex
\section{Methodology}
In this section, we introduce the details of our robust quantity-aware aggregation, named FedRA,
which aims to be quantity-robust and achieve optimal performance.
It contains two core components, i.e. a quantity-robust scorer and a malicious client number estimator.
The quantity-robust scorer calculates scores for clients based on their uploaded updates and quantities.
The malicious client number estimator dynamically determines the number of malicious clients in each round, which is more suitable for the dynamic-ratio setting.
The framework of our FedRA is shown in Figure~\ref{fig:framework}.

\subsection{Quantity-robust Scorer}
\label{sec:qar}
\begin{lemma}
Let $\textbf{g}^i$ and $\textbf{g}^j$ be any pair of independently distributed benign updates and $q^i$ and $q^j$ be the corresponding quantities.
If Assumptions~\ref{assume:sample-sub-exp} and~\ref{assume:iid}  hold, we then have
\begin{equation}
    \begin{aligned}
    \E \,[||\textbf{g}^i - \textbf{g}^j||_1] \leq \sqrt{2\ln 2} \sqrt{\frac{q^i+q^j}{q^iq^j}} ||\boldsymbol \sigma||_1,
    \end{aligned}
\end{equation}
where $\boldsymbol \sigma = [\sigma_1, ... \sigma_d]$, $\sigma_k$ is the standard deviation of the $k$-th dimension of update on the training sample distribution $\mathcal{D}$, and $d$ is the dimension of $\textbf{g}^i$.
\label{lemma:sub-exp-quantity}
\end{lemma}
Lemma~\ref{lemma:sub-exp-quantity} is proved in Appendix~\ref{sec:proof-lemma-sub-exp-quantity}, which reflects that the distance between two benign updates varies according to their quantities.
If the quantities of two clients are larger, the distance between their updates should be smaller.
Inspired by this observation, we design a quantity-robust score, which aims to filter malicious clients considering both their updates and quantities.
The quantity-robust score of the $i$-th client $s(i)$ is defined as follows
\begin{equation}
\begin{aligned}
s(i) = (q^{i})^{\gamma} \sum_{j:i \to j} Q(i, j), \; Q(i, j) =  \sqrt{\frac{q^iq^j}{q^i+q^j}}||\textbf{g}^i - \textbf{g}^j||_1,
\end{aligned}
\end{equation}
where $i \to j$ denotes that $\textbf{g}^j$ belongs to the $(n-\widetilde{m}-2)$ updates closest to $\textbf{g}^i$ in terms of the $Q$ value, $\widetilde{m}$ is the estimated number of malicious client, $\gamma \leqslant 0.5$ is a hyper-parameter.
A smaller quantity-robust score indicates the client is more likely to be benign.

\begin{algorithm}[t]
	\caption{FedRA}
    \label{alg:qra} 
    \everypar={\nl}
		Server initializes $\textbf{w}_{0}$ randomly\;
        \For{each round $t=0,...,T$}
        {Randomly select a client set $\mathcal{C}_{t}$ from $N$ clients\;
        Distribute the global model $\textbf{w}_t$ to selected clients\;
        \For{client $i \in \mathcal{C}_{t}$}{
            $\textbf{g}^i_t, q^i_t \gets \textbf{ClientUpdate}(i, \textbf{w}_t)$\;
            Upload local updates $\textbf{g}^i_t$ and quantity $q^i_t$ to server\;
        }
        \tcp{Quantity-robust scorer}
        $\mathcal{S}\gets \varnothing$, $\widetilde{m} \gets \lceil\frac{n\widetilde{M}}{N}\rceil$\;
        \For{$i = 0, 1,..., n$}{
	       $s(i) \gets (q^i_t)^\gamma \sum_{i \to j} \sqrt{\frac{q^i_tq^j_t}{q^i_t+q^j_t}}||\textbf{g}^i_t - \textbf{g}^j_t||_1$\;
	       $\mathcal{S} \gets \mathcal{S}\cup \{s(i)\}$\;
        }
        Compute the number of clients $c$ to select;
    	\begin{equation*}
    	    c \gets \left\{
            \begin{aligned}
                &n - \widetilde{m} - 1 &\text{fixed-ratio},\\
                &n-\textbf{MCNE}(n, \mathcal{S}) &\text{dynamic-ratio}.\\ 
            \end{aligned}
            \right.
    	\end{equation*}
     Selects $c$ clients $\mathcal{C}$ with smallest scores in $\mathcal{S}$\;
	   $\textbf{g}_t \gets \frac{1}{\sum_{i \in \mathcal{C}}q^i_t}\sum_{i \in \mathcal{C}} q^i_t \textbf{g}^i_t$, $\textbf{w}_t = \textbf{Adam}(\textbf{g}_t)$\;
        }
    \tcp{Malicious client number estimator}
    \SetKwProg{cp}{MCNE($n$, $\mathcal{S})$:}{}{}   
          \cp{}{
           $\mathcal{L} \gets \varnothing$\;
           \For{$i = 0, 1,..., n$}{
                Estimate $\mu_b$, $\sigma_b$, $\mu_m$, $\sigma_m$ by Eq~\ref{eq:est-mu-sigma}\;
                Compute log-likelihood $\hat{l}(i)$ through Eq~\ref{eq:ll-final}\;
                $\mathcal{L} \gets \mathcal{L}\cup\{\hat{l}(i)\}$\;
           }
          $\widetilde{m} \gets \arg\max_{\widetilde{m}}{\mathcal{L}}$\; 
          \KwRet{$\widetilde{m}$}\;}{}
\end{algorithm}

\subsection{Malicious Client Number Estimator}
\label{sec:mce}
Then we need to determine the number of clients to select in each round, which is denoted as $c$.
In the fixed-ratio setting, we can set $c = n - \widetilde{m} -1 = n - \lceil\frac{n\widetilde{M}}{N}\rceil -1$, where $\widetilde{m}$ is the estimated number of malicious clients in each round and $\widetilde{M}$ is the estimated number of overall malicious clients.
However, in the dynamic-ratio setting, the number of malicious clients varies in each round.
Over-estimating $\widetilde{m}$ may result in filtering some benign clients, while underestimating $\widetilde{m}$ may include some malicious updates in the model aggregation.
Therefore, we propose a malicious client number estimator to predict the number of malicious clients in each round.
We treat $\widetilde{m}$ as a parameter of the distribution of quantity-robust scores, and
compute the number of malicious clients $\widetilde{m}$ by maximizing the log-likelihood as follows:
\begin{equation}
    \begin{aligned}
    \widetilde{m} &= \arg \max_{\widetilde{m}} [\ln p(\widetilde{m}, s(1), ... , s(n))] 
    \end{aligned}
\end{equation}

Inspired by ~\citet{reynolds2009gaussian}, we assume the quantity-robust scores are independent and follow two Gaussian distributions.
Meanwhile, as shown in Appendix~\ref{sec:evidence-mcne}, the scores of malicious clients are larger than those of benign clients.
Therefore, we further assume the scores can be separated into two groups by a threshold.
The group of larger scores belongs to malicious clients and follows one of the Gaussian distributions $\mathcal{N}(\mu_m, \sigma_m)$.
The group of the smaller scores belongs to benign clients and follows another Gaussian distribution $\mathcal{N}(\mu_b, \sigma_b)$.
We first sort the scores by ascending order, i.e., $s(i) < s(j), \forall i<j$.
Since $m$ follows the hypergeometric distribution $\mathcal{H}(n, M, N)$, we can compute $\widetilde{m}$ as follows:
\begin{equation}
    \begin{aligned}
    \widetilde{m} &= \arg \max_{\widetilde{m}} [\ln p(\widetilde{m}) + \sum_{i=1}^{n}\ln p(s(i)|\widetilde{m})] \\
    &= \arg \max_{\widetilde{m}} [\ln \tbinom{\widetilde{M}}{\widetilde{m}}\tbinom{N-\widetilde{M}}{n-\widetilde{m}} - (n-\widetilde{m})\ln \sigma_b - \widetilde{m} \ln \sigma_m \\
    &-\sum_{i=1}^{n-\widetilde{m}}\frac{(s(i)-\mu_b)^2}{2\sigma_b^2} - \sum_{i=n-\widetilde{m}+1}^{n}\frac{(s(i)-\mu_m)^2}{2\sigma_m^2}].
    \end{aligned}
    \label{eq:ll-final}
\end{equation}

The mean and variance of the two Gaussian distributions are estimated as follows:
\begin{equation}
    \begin{aligned}
    &\mu_b = \frac{1}{n-\widetilde{m}}\sum_{i=1}^{n-\widetilde{m}}s(i), \sigma_b^2 = \frac{1}{n-\widetilde{m}-1}\sum_{i=1}^{n-\widetilde{m}}(s(i) - \mu_b)^2 \; \\
    &\mu_m = \frac{1}{\widetilde{m}}\sum_{i=n-\widetilde{m}+1}^{n} s(i), \sigma_m^2 = \frac{1}{\widetilde{m}-1}\sum_{i=n-\widetilde{m}+1}^{n}(s(i) - \mu_m)^2.
    \end{aligned}
    \label{eq:est-mu-sigma}
\end{equation}

One may ask how $s(i)$ is computed since it is relevant with $\widetilde{m}$. A reasonable way is to initialize $\widetilde{m} = \lceil\frac{n\widetilde{M}}{N}\rceil$, iteratively compute $s(i)$ and run the malicious client number estimator to update $\widetilde{m}$.
However, our experiments indicate that the performance without iterative approximation is already great enough.

The complete algorithm of our FedRA is shown in Algorithm~\ref{alg:qra}.

%% file: src/theory.tex
\section{Theoretical Analysis}
\subsection{Definitions and Assumptions}
In this subsection, we introduce some definitions and assumptions used in our theoretical analysis.
\begin{definition}
(Sub-exponential random variable).
A random variable $X$ with $\E[X] = \mu$ is called sub-exponential with parameters $(v^2, \alpha)$ if \ $\E[e^{\lambda(X - \mu)}] \leq e^{\frac{1}{2}v^2\lambda^2}$, $\forall |\lambda| < \frac{1}{\alpha}$.
\end{definition}

\begin{definition}
(Lipschitz).
$h$ is $L$-Lipschitz if $|h(\textbf{w}) - h(\textbf{w}')| \leq L||\textbf{w} - \textbf{w}'||_2$, $\forall \ \textbf{w}, \textbf{w}'$.
\end{definition}

\begin{definition}
(Smoothness).
$h$ is $L$-smooth if $||\nabla h(\textbf{w}) - \nabla h(\textbf{w}')|| \\ \leq L||\textbf{w} - \textbf{w}'||_2$, $\forall \ \textbf{w}, \textbf{w}'$.
\end{definition}

\begin{assume}
(Sub-exponential updates).
\label{assume:sample-sub-exp}
For all $\textbf{w}\in \mathcal{W}$, the partial derivative of $f(\cdot;z)$ with respect to the $k$-th dimension of its first argument, denoted as 
$\partial_k f(\textbf{w}; z)$, is sub-exponential with parameters $(v_k^2, \alpha_k)$ 
where $\Var_{z \sim \mathcal{D}}[\partial_k f(\textbf{w}; z)] = \sigma_k^2, 
v_k=\sigma_k$, $\alpha_k<\frac{\sigma_k}{\sqrt{2 \ln 2n}}$, $k \leq d$ and $d$ is the dimension of $\textbf{w}$.
\end{assume}

\begin{assume}
(Smoothness of $F$).
The population loss function $F(\cdot)$ is $L_F$-smooth.
\label{assume:smooth}
\end{assume}

\begin{assume}
(Minimizer in $\mathcal{W}$).
Let $\textbf{w}^* = \arg\min_{\textbf{w} \in \mathcal{W}}F(\textbf{w})$.
We assume that $\nabla F(\textbf{w}^*) = 0$.
\label{assume:minimizer}
\end{assume}

\begin{assume}
(Independent identical distribution).
For any benign client $i$, $z\in D_i$ are independently sampled from  $\mathcal{D}$.
\label{assume:iid}
\end{assume}

Assumptions~\ref{assume:sample-sub-exp},~\ref{assume:smooth}, ~\ref{assume:minimizer} and~\ref{assume:iid} have been applied in previous works~\cite{yin2018byzantine, blanchard2017machine,pmlr-v80-mhamdi18a}.
Then we introduce the definition of quantity-robust.
\begin{lemma}
For any benign update $\textbf{g}^i$ and quantity $q^i$, when Assumptions~\ref{assume:sample-sub-exp} and~\ref{assume:iid} hold, we have 
\begin{equation}
    \begin{aligned}
    \E[\sqrt{q^i}||\textbf{g}^i - \boldsymbol \mu||_1] \leq \sqrt{2\ln 2}||\boldsymbol \sigma||_1,
    \end{aligned}
\end{equation}
where $\boldsymbol \sigma = [\sigma_1, ... \sigma_d]$, $\E_{z \sim \mathcal{D}}[\nabla f(\textbf{w}; z)] = \boldsymbol \mu$.
\label{lemma:low-variance}
\end{lemma}

Lemma~\ref{lemma:low-variance} is proved in Appendix~\ref{sec:proof-low-variance}, which shows the quantity-aware error of benign updates $\E[\sqrt{q^i}||\textbf{g}^i - \boldsymbol \mu||_1]$ is constant.
It inspires us to prove whether a method is quantity-robust by proving the quantity-aware error of every selected update is irrelevant to malicious quantities.
The formulation of the quantity-robust definition is as follows:
\begin{definition}
(Quantity-robust).
An aggregation rule is quantity-robust, only when every selected update $\textbf{g}$ and quantity $q$ satisfies $\E[\sqrt{q}||\textbf{g} - \boldsymbol \mu||_1] \leq C$, where $C$ is irrelevant to malicious quantities and updates.

\end{definition}

\subsection{FedRA is Quantity-robust}

\begin{prop}
Let $\textbf{g}^1$, ..., $\textbf{g}^{n-m}$ be independent benign updates with quantities $q^1$, ..., $q^{n-m}$, where $\textbf{g}^i \in \mathbb{R}^d$, and $\E\,[\textbf{g}^i] = \boldsymbol \mu$.
Let $\textbf{g}^{n-m+1}$, ..., $\textbf{g}^{n}$ be  malicious updates in $\mathbb{R}^d$ and $q^{n-m+1}$, ..., $q^{n}$ be their quantities.
Suppose that Assumptions~\ref{assume:sample-sub-exp} and~\ref{assume:iid} hold for all benign updates.
If $2m+2\leqslant n$, $\gamma = 0.5$, then FedRA satisfies 
\begin{equation}
\E[\sqrt{q}||\textbf{g} - \overline{\textbf{g}}||_1] \leqslant A_{n,m} B_q ||\boldsymbol \sigma||_1,
\end{equation}
where $\textbf{g}$ and $q$ is the quantity and update of the client with the smallest quantity-robust score,
\begin{equation}
    \begin{aligned}
    A_{n,m} &= \sqrt{2\ln 2}(n-m) + m\sqrt{2 \ln 2(n-m)}, \\
    B_q &= \frac{\max_{\text{benign $i$}}[q^i + 1] \min_{\text{benign $i$}}}{\sum_{i \in \mathcal{B}}q^i}, \\
    \overline{\textbf{g}} &= \frac{1}{\displaystyle\sum_{j:i_* \to j \atop \text{benign $j$}}q^j}\sum_{j:i_* \to j \atop \text{benign $j$}}q^j\textbf{g}^j, 
    \end{aligned}
\end{equation}
$\mathcal{B}$ is denoted as the (n-m-2) benign clients with smallest quantities.
If $\displaystyle\max_{\text{malicious i}}q^i < \sum_{i \in \mathcal{B}}q^i$, then $\textbf{g}$ satisfies
\begin{equation}
    \begin{aligned}
        \E[\sqrt{q}||\textbf{g} - \boldsymbol \mu||_1] \leqslant  (A_{n,m}B_q + C_{n,m})||\boldsymbol \sigma||_1,
    \end{aligned}
\end{equation}
where $C_{n,m} = \sqrt{2 \ln 2(n-m)} + \sqrt{2 \ln 2}$.
\label{prop:fedra}
\end{prop}

\begin{remark}
Proposition~\ref{prop:fedra} can be extended to other selected updates and quantities with larger $B_q$ irrelevant to malicious quantities and smaller benign client set $\mathcal{B}$.
\end{remark}

Proposition~\ref{prop:fedra} is proved in Appendix~\ref{sec:proof-fedra}.
Since the error of FedRA is controlled by $A_{n,m}$, $B_q$, $||\boldsymbol \sigma||_1$, and $C_{n, m}$ which are irrelevant to  malicious quantities and updates,  FedRA is quantity-robust.

We state the statistical error guarantees of FedRA for smooth non-convex $F$.
\begin{theorem}
Assume that Assumptions~\ref{assume:sample-sub-exp}, ~\ref{assume:smooth} ~\ref{assume:minimizer} and~\ref{assume:iid} hold, $2m + 2 < n$, $\gamma = 0.5$ and $\displaystyle\max_{\text{malicious i}}q^i < \sum_{i \in \mathcal{B}}q^i$.
Choose $\eta = 1/L_F$.
If $\textbf{w}_{t+1} = \textbf{w}_{t} - \eta \textbf{g}_t$, after $T$ iterations with FedRA, we have 
\begin{equation}
    \begin{aligned}
    \min_{t=0,...,T}\E[||\nabla F(\textbf{w}_t)||_2^2] \leq \frac{2L_F}{T}\E [F(\textbf{w}_{0}) - F(\textbf{w}^*)] \\
    + \frac{1}{T}\sum_{t=0}^{T-1} ((A_{n,m}B_q + C_{n,m})||\boldsymbol \sigma||_1)^2.
    \end{aligned}
\end{equation}
\label{theorem:non-convex-fedra}
\end{theorem}

Theorem~\ref{theorem:non-convex-fedra} is proved in Appendix~\ref{sec:proof-non-convex-fedra}.

%% file: src/exp.tex
\begin{table}[!t]
\centering
\caption{Dataset statistics.}
\scalebox{0.83}{
\begin{tabular}{cccccccc}
\Xhline{1.5pt}
\multirow{2}{*}{Dataset} & \multirow{2}{*}{\#Classes} & \multirow{2}{*}{\#Train} & \multirow{2}{*}{\#Test} & \multirow{2}{*}{\#Clients} & \multicolumn{3}{c}{\#Train per client} \\
                         &                                      &                          &                         &                            & Mean        & Std        & Max       \\ \hline
MNIST                      & 10                         &    60,000                &    10,000               &     3,025                  &    19.77      &   179.28   &   6,820   \\
CIFAR10                    & 10                         &    50,000                &    10,000               &     3,115                  &    16.05      &   200.14   &   8,933   \\
MIND                        & 18                         &    71,068                &    20,307               &     2,880                  &    24.68      &   299.61   &   9,398   \\ \Xhline{1.5pt}
\end{tabular}
}
\label{tab:data}
\end{table}

\begin{table*}[!t]
\centering
\caption{Performance of different methods in IID fixed-ratio settings, with the best results in \textbf{bold} and second-best ones in \ul{underlined}. *Best QI refers to the best quantity-ignorant method.}
\scalebox{0.88}{
\begin{tabular}{c|c|ccccc|ccccc|ccccc}
\Xhline{1.5pt}
\multirow{2}{*}{Dataset} & \multirow{2}{*}{Method} & \multicolumn{5}{c|}{Label Flip}                                                    & \multicolumn{5}{c|}{LIE}                                                           & \multicolumn{5}{c}{Optimize}                                                       \\ \cline{3-17} 
                         &                         & 0              & 1              & 2              & 5              & 10             & 0              & 1              & 2              & 5              & 10             & 0              & 1              & 2              & 5              & 10             \\ \hline
\multirow{6}{*}{MNIST}   & Best QI*                 & 98.77          & 98.77          & 98.77          & \ul{98.77}    & \ul{98.77}    & 94.50          & 94.50          & \ul{94.50}    & \ul{94.50}    & \ul{94.50}    & \ul{98.76}    & \ul{98.76}    & \ul{98.76}    & \ul{98.76}    & \ul{98.76}    \\
                         & FedAdam                 & 98.68          & 97.77          & 97.05          & 94.42          & 90.38          & 92.52          & 38.23          & 22.63          & 23.45          & 16.09          & 9.82           & 9.46           & 8.92           & 8.92           & 8.92           \\
                         & Norm-Bound              & 99.08          & 98.65          & 98.33          & 97.47          & 96.09          & 98.25          & \ul{96.31}    & 94.02          & 42.81          & 12.78          & 97.59          & 97.57          & 95.80          & 8.92           & 8.92           \\
                         & RFA                     & \textbf{99.31} & \ul{99.11}    & \ul{98.78}    & 98.36          & 97.59          & \ul{99.05}    & 90.28          & 58.91          & 25.61          & 19.24          & 97.77          & 10.68          & 8.92           & 8.92           & 8.92           \\
                         & Truncate                & 97.20          & 97.19          & 97.18          & 97.16          & 97.17          & 93.83          & 93.28          & 93.31          & 93.28          & 93.25          & 52.87          & 52.44          & 56.61          & 53.27          & 55.22          \\
                         & FedRA                   & \ul{99.30}    & \textbf{99.25} & \textbf{99.30} & \textbf{99.25} & \textbf{99.31} & \textbf{99.20} & \textbf{99.16} & \textbf{99.21} & \textbf{99.08} & \textbf{99.20} & \textbf{99.21} & \textbf{99.27} & \textbf{99.22} & \textbf{99.25} & \textbf{99.26} \\ \hline
\multirow{6}{*}{CIFAR10} & Best QI*                 & 51.42          & 51.42          & \ul{51.42}    & \ul{51.42}    & \ul{51.42}    & \ul{50.65}    & \ul{50.65}    & \ul{50.65}    & \ul{50.65}    & \ul{50.65}    & \ul{52.60}    & \ul{52.60}    & \ul{52.60}    & \ul{52.60}    & \ul{52.60}    \\
                         & FedAdam                 & 56.95          & 38.60          & 31.32          & 19.35          & 15.02          & 13.43          & 10.12          & 10.18          & 10.01          & 10.00          & 12.64          & 10.00          & 10.00          & 10.00          & 10.00          \\
                         & Norm-Bound              & 60.22          & 43.81          & 35.44          & 27.44          & 17.54          & 20.27          & 13.61          & 10.62          & 13.74          & 12.21          & 25.13          & 10.00          & 10.00          & 10.00          & 10.00          \\
                         & RFA                     & \textbf{62.19} & \ul{52.11}    & 46.01          & 36.86          & 29.34          & 19.62          & 10.03          & 10.01          & 10.00          & 10.08          & 12.87          & 10.00          & 10.00          & 10.00          & 10.00          \\
                         & Truncate                & 47.45          & 45.81          & 45.57          & 45.57          & 45.57          & 13.46          & 14.23          & 11.25          & 11.25          & 11.25          & 12.68          & 12.37          & 12.41          & 12.41          & 12.41          \\
                         & FedRA                   & \ul{60.44}    & \textbf{61.73} & \textbf{62.11} & \textbf{62.28} & \textbf{62.37} & \textbf{62.58} & \textbf{62.58} & \textbf{62.58} & \textbf{62.58} & \textbf{62.58} & \textbf{62.69} & \textbf{62.69} & \textbf{62.69} & \textbf{62.69} & \textbf{62.69} \\ \hline
\multirow{6}{*}{MIND}    & Best QI*                 & 66.52          & 66.52          & 66.52          & 66.52          & 66.52          & 61.19          & \ul{61.19}    & \ul{61.19}    & \ul{61.19}    & \ul{61.19}    & \ul{67.25}    & \ul{67.25}    & \ul{67.25}    & \ul{67.25}    & \ul{67.25}    \\
                         & FedAdam                 & 67.03          & 63.35          & 62.03          & 60.91          & 59.01          & 53.69          & 31.84          & 31.84          & 4.83           & 4.83           & 28.70          & 10.73          & 9.77           & 8.26           & 8.39           \\
                         & Norm-Bound              & 67.66          & 63.30          & 62.19          & 61.02          & 59.35          & 54.41          & 31.84          & 31.84          & 4.83           & 4.83           & 54.88          & 23.35          & 7.52           & 6.90           & 6.97           \\
                         & RFA                     & \textbf{70.30} & 66.49          & 65.47          & 63.76          & 61.34          & \ul{64.52}    & 31.84          & 31.84          & 31.84          & 4.83           & 50.37          & 9.04           & 6.35           & 5.45           & 4.98           \\
                         & Truncate                & 67.92          & \ul{67.68}    & \ul{67.68}    & \ul{67.68}    & \ul{67.68}    & 53.64          & 52.70          & 52.70          & 53.11          & 53.11          & 54.76          & 54.59          & 54.59          & 54.59          & 54.59          \\
                         & FedRA                   & \ul{69.96}    & \textbf{69.61} & \textbf{69.94} & \textbf{70.15} & \textbf{70.27} & \textbf{70.71} & \textbf{70.73} & \textbf{70.70} & \textbf{70.70} & \textbf{70.70} & \textbf{70.85} & \textbf{70.85} & \textbf{70.85} & \textbf{70.85} & \textbf{70.85} \\ \Xhline{1.5pt}
\end{tabular}
}
\label{tab:fixed-iid}
\end{table*}

\begin{table}[!t]
\centering
\caption{Performance of quantity-ignorant methods in IID fixed-ratio settings.}
\scalebox{0.87}{
\begin{tabular}{c|c|ccccc}
\Xhline{1.5pt}
Dataset                  & Attack     & Krum  & Median & Trimean & mKrum & Bulyan \\ \hline
\multirow{3}{*}{MNIST}   & Label Flip & 87.01 & 88.62  & 96.93   & 98.77 & 96.89  \\
                         & LIE        & 84.11 & 85.24  & 91.55   & 89.51 & 94.50  \\
                         & Optimize   & 87.04 & 46.35  & 30.34   & 98.76 & 96.92  \\ \hline
\multirow{3}{*}{CIFAR10} & Label Flip & 51.42 & 27.85  & 49.39   & 50.37 & 47.18  \\
                         & LIE        & 50.65 & 18.53  & 15.55   & 22.68 & 24.58  \\
                         & Optimize   & 52.60 & 19.04  & 18.77   & 51.69 & 51.48  \\ \hline
\multirow{3}{*}{MIND}    & Label Flip & 64.40 & 56.85  & 65.11   & 65.54 & 66.52  \\
                         & LIE        & 61.19 & 55.37  & 54.39   & 56.25 & 59.57  \\
                         & Optimize   & 62.53 & 54.10  & 38.94   & 67.25 & 67.11  \\
                         \Xhline{1.5pt}
\end{tabular}
}
\label{tab:fixed-iid-ign}
\end{table}

\begin{table*}[!t]
\caption{Performance of different methods in IID dynamic-ratio settings, with the best results in \textbf{bold} and second-best ones in \ul{underlined}. *Best QI refers to the best quantity-ignorant method.}
\scalebox{0.88}{
\begin{tabular}{c|c|ccccc|ccccc|ccccc}
\Xhline{1.5pt}
\multirow{2}{*}{Dataset} & \multirow{2}{*}{Method} & \multicolumn{5}{c|}{Label Flip}                                                    & \multicolumn{5}{c|}{LIE}                                                           & \multicolumn{5}{c}{Optimize}                                                       \\ \cline{3-17} 
                         &                         & 0              & 1              & 2              & 5              & 10             & 0              & 1              & 2              & 5              & 10             & 0              & 1              & 2              & 5              & 10             \\ \hline
\multirow{6}{*}{MNIST}   & Best QI                 & 98.65     & 98.65      & \ul{98.65}  & \textbf{98.65}    & \ul{98.65}    & 93.16          & 93.16          & 93.16          & \ul{93.16}    & \ul{93.16}    & 97.11          & 97.11          & \ul{97.11}    & \ul{97.11}    & \ul{97.11}    \\
                         & FedAdam                 & 98.74          & 97.84          & 97.17          & 95.21          & 92.89          & 93.25          & 50.42          & 24.25          & 16.69          & 16.26          & 9.69           & 9.67           & 8.92           & 8.92           & 8.92           \\
                         & Norm-Bound              & \ul{99.09}          & \ul{98.75}          & 98.51          & 97.76          & 96.56          & 98.35          & \ul{96.49}    & \ul{94.32}    & 76.15          & 19.67          & \ul{97.76}    & \ul{97.56}    & 96.29          & 8.92           & 8.92           \\
                         & RFA                     & \textbf{99.31} & \textbf{99.08}    & \textbf{98.79}    & 98.38          & 97.54          & \ul{98.70}    & 91.21          & 64.81          & 28.63          & 18.72          & 96.32          & 9.79           & 8.92           & 8.92           & 8.92           \\
                         & Truncate                & 97.23          & 97.00          & 96.98          & 96.92          & 96.93          & 84.71          & 76.88          & 74.34          & 73.15          & 72.85          & 9.92           & 9.78           & 9.78           & 9.97           & 10.02          \\
                         & FedRA                   & 98.53    & 98.46 &  98.48 & \ul{98.49} & \textbf{98.66} & \textbf{99.27} & \textbf{99.32} & \textbf{99.30} & \textbf{99.31} & \textbf{99.28} & \textbf{98.87} & \textbf{98.95} & \textbf{98.93} & \textbf{98.81} & \textbf{98.90} \\ \hline
\multirow{6}{*}{CIFAR10} & Best QI                 & 51.75          & \ul{51.75}    & \ul{51.75}    & \ul{51.75}    & \ul{51.75}    & \ul{52.14}    & \ul{52.14}    & \ul{52.14}    & \ul{52.14}    & \ul{52.14}    & \ul{52.85}    & \ul{52.85}    & \ul{52.85}    & \ul{52.85}    & \ul{52.85}    \\
                         & FedAdam                 & 56.30          & 40.93          & 34.89          & 22.43          & 15.86          & 13.61          & 10.68          & 10.40          & 10.00          & 10.32          & 11.76          & 10.00          & 10.00          & 10.00          & 10.00          \\
                         & Norm-Bound              & \ul{60.14}    & 45.25          & 38.83          & 27.02          & 19.41          & 22.95          & 13.58          & 11.20          & 10.59          & 13.59          & 26.79          & 10.00          & 10.00          & 10.00          & 10.00          \\
                         & RFA                     & 62.31          & 50.36          & 45.98          & 33.32          & 22.42          & 18.56          & 10.01          & 10.16          & 10.02          & 10.10          & 15.90          & 10.00          & 10.00          & 10.00          & 10.00          \\
                         & Truncate                & 48.01          & 41.82          & 45.28          & 45.05          & 44.11          & 14.38          & 13.49          & 13.76          & 13.71          & 13.62          & 13.89          & 10.67          & 12.17          & 13.42          & 12.94          \\
                         & FedRA                   & \textbf{61.86} & \textbf{61.32} & \textbf{62.68} & \textbf{62.59} & \textbf{63.92} & \textbf{64.58} & \textbf{64.80} & \textbf{64.53} & \textbf{64.22} & \textbf{64.23} & \textbf{64.32} & \textbf{64.64} & \textbf{64.26} & \textbf{64.45} & \textbf{64.21} \\ \hline
\multirow{6}{*}{MIND}    & Best QI                 & 66.46          & 66.46          & 66.46          & 66.46          & 66.46          & \ul{59.09}    & \ul{59.09}    & \ul{59.09}    & \ul{59.09}    & \ul{59.09}    & \ul{66.50}    & \ul{66.50}    & \ul{66.50}    & \ul{66.50}    & \ul{66.50}    \\
                         & FedAdam                 & 65.32          & 63.23          & 61.34          & 60.96          & 59.16          & 53.74          & 31.84          & 31.79          & 31.84          & 4.83           & 17.53          & 12.12          & 10.04          & 9.53           & 8.82           \\
                         & Norm-Bound              & 66.13          & 63.54          & 61.24          & 60.61          & 58.99          & 54.18          & 31.84          & 20.02          & 31.84          & 4.83           & 55.65          & 11.92          & 8.04           & 5.97           & 5.57           \\
                         & RFA                     & 67.53          & 65.57          & 65.31          & 62.12          & 59.81          & 61.69          & 31.82          & 31.84          & 31.84          & 31.84          & 41.09          & 8.90           & 6.03           & 6.50           & 1.74           \\
                         & Truncate                & \ul{68.49}    & \ul{67.93}    & \ul{67.83}    & \ul{67.90}    & \ul{67.84}    & 51.94          & 46.68          & 46.57          & 46.52          & 46.51          & 23.64          & 17.16          & 17.32          & 17.47          & 17.47          \\
                         & FedRA                   & \textbf{69.96} & \textbf{69.92} & \textbf{69.97} & \textbf{70.11} & \textbf{70.49} & \textbf{71.09} & \textbf{71.12} & \textbf{71.11} & \textbf{71.11} & \textbf{71.11} & \textbf{71.02} & \textbf{70.89} & \textbf{71.05} & \textbf{71.13} & \textbf{71.12} \\ \Xhline{1.5pt}
\end{tabular}
}
\label{tab:dyn-iid}
\end{table*}

\begin{table}[!t]
\centering
\caption{Performance of quantity-ignorant methods in IID dynamic-ratio settings.}
\scalebox{0.87}{
\begin{tabular}{c|c|ccccc}
\Xhline{1.5pt}
Dataset                  & Attack     & Krum  & Median & Trimean & mKrum & Bulyan \\ \hline
\multirow{3}{*}{MNIST}   & Label Flip & 87.67 & 87.90  & 96.86   & 98.65 & 97.30  \\
                         & LIE        & 83.03 & 87.05  & 79.44   & 93.16 & 79.08  \\
                         & Optimize   & 86.03 & 44.92  & 10.70   & 97.11 & 96.59  \\ \hline
\multirow{3}{*}{CIFAR10} & Label Flip & 51.75 & 33.25  & 49.39   & 51.71 & 49.61  \\
                         & LIE        & 52.14 & 18.23  & 13.31   & 21.44 & 17.63  \\
                         & Optimize   & 52.56 & 12.63  & 16.60   & 52.85 & 52.16  \\ \hline
\multirow{3}{*}{MIND}    & Label Flip & 61.92 & 56.38  & 65.31   & 65.62 & 66.46  \\
                         & LIE        & 59.09 & 55.15  & 54.11   & 57.46 & 55.38  \\
                         & Optimize   & 62.81 & 54.12  & 38.18   & 65.37 & 66.50  \\ \Xhline{1.5pt}
\end{tabular}
}
\label{tab:dyn-iid-ign}
\end{table}
\section{Experiments}
\label{sec:exp}
\subsection{Datasets and Experimental Settings}
\textbf{Dataset.}
We conduct experiments on three public datasets:, i.e., MNIST~\citep{lecun1998gradient}, CIFAR10~\citep{krizhevsky2009learning},  and MIND~\citep{wu-etal-2020-mind}.
The quantities follow a long-tailed distribution, i.e., log-normal distribution.
The average sample size of clients is around 20, and the $\sigma$ of the log-normal distribution is 3.
We randomly shuffle the dataset and partition it according to the quantities.
For the IID setting, we randomly divide the dataset into clients' local datasets.
For the non-IID setting, we guarantee the local datasets of most clients contain only one class.
The detailed statistics of the datasets are shown in Table~\ref{tab:data}.

\textbf{Configurations.}
In our experiments, we use CNN networks as base models for the MNIST and CIFAR10 datasets.
For the MIND dataset, we use a Text-CNN as the base model, and initialize the word embedding matrix with pre-trained Glove embeddings~\citep{pennington2014glove}.
We apply FedAdam~\citep{reddi2021adaptive} to accelerate model convergence in all methods.
We apply dropout with dropout rate 0.2 to mitigate over-fitting.
The learning rate is 0.001 for CIFAR10, and 0.0001 for MNIST and MIND.
The maximum of training rounds is 10,000 for MNIST and CIFAR10, and 15,000 for MIND.
The ratio of malicious clients $M/N$ is 0.1.
The number of clients sampled in each round $n$ is 50. $\gamma$ is 0.1.
The server estimates $\widetilde{M}$ as $M$.

\begin{table*}[!t]
\caption{Performance of different methods in non--IID dynamic-ratio settings, with the best results in \textbf{bold} and second-best ones in \ul{underlined}. *Best QI refers to the best quantity-ignorant method.}
\scalebox{0.88}{
\begin{tabular}{c|c|ccccc|ccccc|ccccc}
\Xhline{1.5pt}
\multirow{2}{*}{Dataset} & \multirow{2}{*}{Method} & \multicolumn{5}{c|}{Label Flip}                                                    & \multicolumn{5}{c|}{LIE}                                                           & \multicolumn{5}{c}{Optimize}                                                       \\ \cline{3-17} 
                         &                         & 0              & 1              & 2              & 5              & 10             & 0              & 1              & 2              & 5              & 10             & 0              & 1              & 2              & 5              & 10             \\ \hline
\multirow{6}{*}{MNIST}   & Best QI                 & 98.49          & 98.49          & 98.49          & \textbf{98.49} & \textbf{98.49} & 89.87          & \ul{89.87}    & \ul{89.87}    & \ul{89.87}    & \ul{89.87}    & \ul{96.94}    & \ul{96.94}    & \ul{96.94}    & \ul{96.94}    & \ul{96.94}    \\
                         & FedAdam                 & 98.67          & 97.55          & 96.78          & 93.42          & 87.58          & 90.59          & 44.44          & 28.09          & 15.30          & 19.13          & 10.36          & 11.39          & 11.35          & 11.35          & 11.35          \\
                         & Norm-Bound              & \textbf{99.03} & \ul{98.69}    & \ul{98.54}    & 97.86          & 95.81          & 97.61          & 83.45          & 69.37          & 36.57          & 19.98          & 95.92          & 85.61          & 63.39          & 11.35          & 11.35          \\
                         & RFA                     & \ul{98.99}    & \textbf{98.93} & \textbf{98.54} & \ul{97.89}    & \ul{95.78}    & \ul{98.46}    & 72.32          & 53.56          & 26.02          & 19.46          & 95.62          & 8.56           & 11.63          & 11.35          & 11.35          \\
                         & Truncate                & 88.88          & 87.76          & 86.95          & 87.29          & 87.28          & 51.37          & 45.40          & 45.24          & 44.57          & 44.93          & 11.24          & 13.05          & 13.32          & 13.34          & 13.34          \\
                         & FedRA                   & 95.28          & 95.68          & 95.16          & 95.65          & 95.31          & \textbf{99.00} & \textbf{99.03} & \textbf{99.04} & \textbf{99.05} & \textbf{99.04} & \textbf{98.48} & \textbf{98.47} & \textbf{98.49} & \textbf{98.58} & \textbf{98.60} \\ \hline
\multirow{6}{*}{CIFAR10} & Best QI                 & 49.88          & \textbf{49.88} & \textbf{49.88} & \textbf{49.88} & \textbf{49.88} & \ul{20.18}    & \ul{20.18}    & \ul{20.18}    & \ul{20.18}    & \ul{20.18}    & \ul{40.90}    & \ul{40.90}    & \ul{40.90}    & \ul{40.90}    & \ul{40.90}    \\
                         & FedAdam                 & 50.70          & 34.62          & 29.33          & 17.06          & 14.20          & 14.07          & 10.24          & 10.17          & 10.01          & 10.06          & 14.97          & 10.00          & 10.00          & 10.00          & 10.00          \\
                         & Norm-Bound              & \textbf{53.58} & 37.97          & 31.03          & 22.97          & 16.58          & 14.34          & 10.65          & 10.69          & 11.74          & 10.54          & 17.10          & 10.00          & 10.00          & 10.00          & 10.00          \\
                         & RFA                     & \ul{51.59}    & 38.11          & 30.00          & 21.10          & 17.89          & 13.86          & 10.00          & 10.05          & 10.06          & 10.01          & 14.17          & 10.00          & 10.00          & 10.00          & 10.00          \\
                         & Truncate                & 22.67          & 25.16          & 26.91          & 24.92          & 23.99          & 13.97          & 13.74          & 13.77          & 14.41          & 13.98          & 13.50          & 10.96          & 10.08          & 11.42          & 12.88          \\
                         & FedRA                   & 44.29          & \ul{45.90}    & \ul{45.57}    & \ul{47.93}    & \ul{49.69}    & \textbf{59.48} & \textbf{59.45} & \textbf{59.49} & \textbf{59.04} & \textbf{59.36} & \textbf{59.01} & \textbf{58.92} & \textbf{59.25} & \textbf{58.38} & \textbf{58.42} \\ \hline
\multirow{6}{*}{MIND}    & Best QI                 & 62.59          & 62.59          & \ul{62.59}    & \ul{62.59}    & \ul{62.59}    & \ul{59.65}    & \ul{59.65}    & \ul{59.65}    & \ul{59.65}    & \ul{59.65}    & \ul{60.06}    & \ul{60.06}    & \ul{60.06}    & \ul{60.06}    & \ul{60.06}    \\
                         & FedAdam                 & 63.60          & 62.01          & 60.93          & 57.63          & 55.19          & 54.96          & 31.84          & 4.52           & 4.52           & 4.52           & 26.93          & 13.81          & 11.48          & 7.66           & 1.74           \\
                         & Norm-Bound              & 65.92          & 62.91          & 60.79          & 58.64          & 56.36          & 54.81          & 31.84          & 31.84          & 31.84          & 31.84          & 4.61           & 17.36          & 14.00          & 5.96           & 1.36           \\
                         & RFA                     & \textbf{68.10} & \ul{65.37}    & 61.86          & 55.88          & 55.62          & 56.53 & 31.84          & 31.84          & 31.84          & 4.52           & 46.73          & 15.49          & 8.94           & 1.53           & 0.96           \\
                         & Truncate                & 62.46          & 61.59          & 62.04          & 61.52          & 61.57          & 31.84          & 31.84          & 31.84          & 31.93          & 29.65          & 29.64          & 25.50          & 26.22          & 21.31          & 24.47          \\
                         & FedRA                   & \ul{66.33}    & \textbf{67.07} & \textbf{66.30} & \textbf{67.02} & \textbf{66.86} & \textbf{69.20} & \textbf{69.17} & \textbf{69.22} & \textbf{69.25} & \textbf{69.21} & \textbf{69.20} & \textbf{69.22} & \textbf{69.04} & \textbf{69.16} & \textbf{69.17} \\ \Xhline{1.5pt}
\end{tabular}
}
\label{tab:non-iid}
\end{table*}

\begin{table}[!t]
\caption{Performance of quantity-ignorant methods in non-IID dynamic-ratio settings.}
\scalebox{0.87}{
\begin{tabular}{c|c|ccccc}
\Xhline{1.5pt}
Dataset                  & Attack     & Krum  & Median & Trimean & mKrum & Bulyan \\ \hline
\multirow{3}{*}{MNIST}   & Label Flip & 36.71 & 39.25  & 88.13   & 98.49 & 93.90  \\
                         & LIE        & 36.12 & 33.62  & 50.93   & 89.87 & 47.63  \\
                         & Optimize   & 35.36 & 9.67   & 10.99   & 96.94 & 83.17  \\ \hline
\multirow{3}{*}{CIFAR10} & Label Flip & 21.51 & 20.15  & 32.58   & 49.88 & 20.68  \\
                         & LIE        & 20.18 & 13.40  & 13.62   & 16.64 & 13.60  \\
                         & Optimize   & 21.91 & 14.81  & 13.27   & 40.90 & 25.22  \\ \hline
\multirow{3}{*}{MIND}    & Label Flip & 50.89 & 56.04  & 59.93   & 62.59 & 60.45  \\
                         & LIE        & 54.07 & 35.32  & 31.33   & 46.29 & 59.65  \\
                         & Optimize   & 53.59 & 17.59  & 10.11   & 56.01 & 60.06  \\ \Xhline{1.5pt}
\end{tabular}
}
\label{tab:non-iid-ign}
\end{table}

\textbf{Baselines.}
We compare our FedRA with several baseline methods, including
1) \textbf{Median}~\citep{yin2018byzantine}, applying coordinate-wise median on each dimension of updates;
2) \textbf{Trimean}~\citep{yin2018byzantine}, applying coordinate-wise trimmed-mean on each dimension of updates;
3) \textbf{Krum}~\citep{blanchard2017machine}, selecting the update that is closest to a subset of neighboring updates based on the square distance;
4) \textbf{mKrum}~\citep{blanchard2017machine}, a variance of Krum that selects multiple updates and averages the selected updates;
5) \textbf{Bulyan}~\citep{pmlr-v80-mhamdi18a}, selecting multiple clients with mKrum and aggregating the selected updates with Trimean;
6) \textbf{Norm-bounding}~\citep{sun2019can}, clipping the $L_2$ norm of each update with a certain threshold;
7) \textbf{RFA}~\citep{pillutla2019robust}, applying an approximation algorithm to minimize the geometric median of updates;
8) \textbf{Truncate}~\citep{portnoy2020towards}, limiting the quantity of each client under a dynamic threshold in each round and applying quantity-aware Trimean.

\textbf{Attack Model.}
We suppose an attacker controls malicious clients. Each malicious client, if sampled, submits malicious updates and a malicious quantity.
We implement three existing untargeted poisoning attack methods to create malicious updates, including
1) \textbf{Label Flip}~\cite{10.5555/3489212.3489304}: a data poisoning attack that manipulates labels of training samples;
2) \textbf{LIE}~\cite{NEURIPS2019_ec1c5914}: adding small enough noise in the mean of benign updates to circumvent defenses;
3) \textbf{Optimize}~\cite{10.5555/3489212.3489304}: a model poisoning attack that adds noise in the opposite position of benign updates.
To create malicious quantities, 
we set the malicious quantity $\alpha_q \in \{0, 1, 2, 5, 10\}$.
It is noted that when $\alpha_q$ equals or is less than 10, the malicious quantity is still smaller than the maximum benign quantity in training dataset.
Therefore, the malicious quantity is in a reasonable range in our settings.

\subsection{Performance in IID Fixed-ratio Setting}
In this subsection, we conduct experiments in the fixed-ratio setting to analyze the effectiveness of our quantity-robust scorer.
The experimental results are shown in Table~\ref{tab:fixed-iid} and Table~\ref{tab:fixed-iid-ign}.
We can make the following observations from the tables.
First, our FedRA outperforms the best quantity-ignorant defense in the fixed-ratio settings in the vast majority of cases.
This is because our method performs weighted averaging on selected updates based on their quantities.
Second, our FedRA has stable performance with different quantity-enlarging factors and outperforms other quantity-aware methods.
This is because FedRA can defend against quantity-enhanced attacks by jointly considering updates and quantities to filter malicious clients.
These two observations reflect the effectiveness of our FedRA algorithm.
Third, the performance of quantity-aware defenses, i.e., RFA and Norm-bound, becomes worse with larger quantity-enlarging factors.
This is because these quantity-aware defenses by default treat received quantities as benign, which is vulnerable to quantity-enhanced attacks.
Finally, Truncate has stable performance with different quantity-enlarging factors, but its performance is lower than FedRA.
This is because the Truncate algorithm is quantity-robust by limiting quantities submitted by malicious clients.
However, it also restricts quantities of benign clients.
Meanwhile, it does not filter malicious clients by jointly considering quantities and updates.
Thus, it has sub-optimal performance.

\subsection{Performance in IID Dynamic-ratio Setting}
\label{sec:dyn-non-iid}
In this subsection, we conduct experiments in the IID dynamic-ratio setting to analyze the effectiveness of our malicious client number selector.
The experimental results are shown in Table~\ref{tab:dyn-iid} and Table~\ref{tab:dyn-iid-ign}.
Besides the same observations in the fixed-ratio setting, we can make several additional observations.
First, our FedRA has stable performance with different quantity-enlarging factors.
It outperforms or has similar performance as the best quantity-ignorant defense.
This shows the effectiveness of our FedRA with the malicious client estimator.
Second, the algorithms that need to estimate the number or the upper bound of malicious clients, i.e., mKrum, Trimean, Bulyan, and Truncate, have lower performance in the dynamic-ratio setting than in the fixed-ratio setting.
This is because the number of malicious clients in each round changes dynamically.
Over-estimating makes a subset of benign clients excluded, while under-estimating causes some malicious clients selected in some rounds.

\subsection{Performance in non-IID Setting}
In this subsection, we analyze the performance of our FedRA in the non-IID dynamic-ratio setting.
The experimental results are shown in Table~\ref{tab:non-iid} and Table~\ref{tab:non-iid-ign}, where we can have several observations.
First, our FedRA consistently outperforms both quantity-ignorant and quantity-aware methods under LIE-based and Optimize-based quantity-enhanced attacks.
This is because we apply quantity-robust scorer which helps accurately filter out malicious clients and effectively aggregate updates with quantities.
Second, the performance of our FedRA drops under Label-flip-based attacks.
Suppose the sample sizes of all classes are the same.  
If a benign client has a large dataset with a single class, then it is likely there are fewer benign clients holding the dataset with the same class since in the non-IID settings most clients only have datasets with a single class.
Therefore, benign clients with large quantities are likely to be filtered out, especially under less advanced attacks like Label flip.
Finally, compared to the performance under IID settings, the performance of all methods under non-IID settings drops.
This is because the non-IID settings introduce a drift in local updates, leading to slow model convergence~\cite{karimireddy2020scaffold,li2020federated}.
Meanwhile, in non-IID settings the variance of clients' local updates becomes larger, giving malicious updates more chances to circumvent defenses.

\subsection{Ablation Study}

 \begin{figure}[!t]
     \centering
     \subfigure[CIFAR10]{\includegraphics[width=0.23\textwidth]{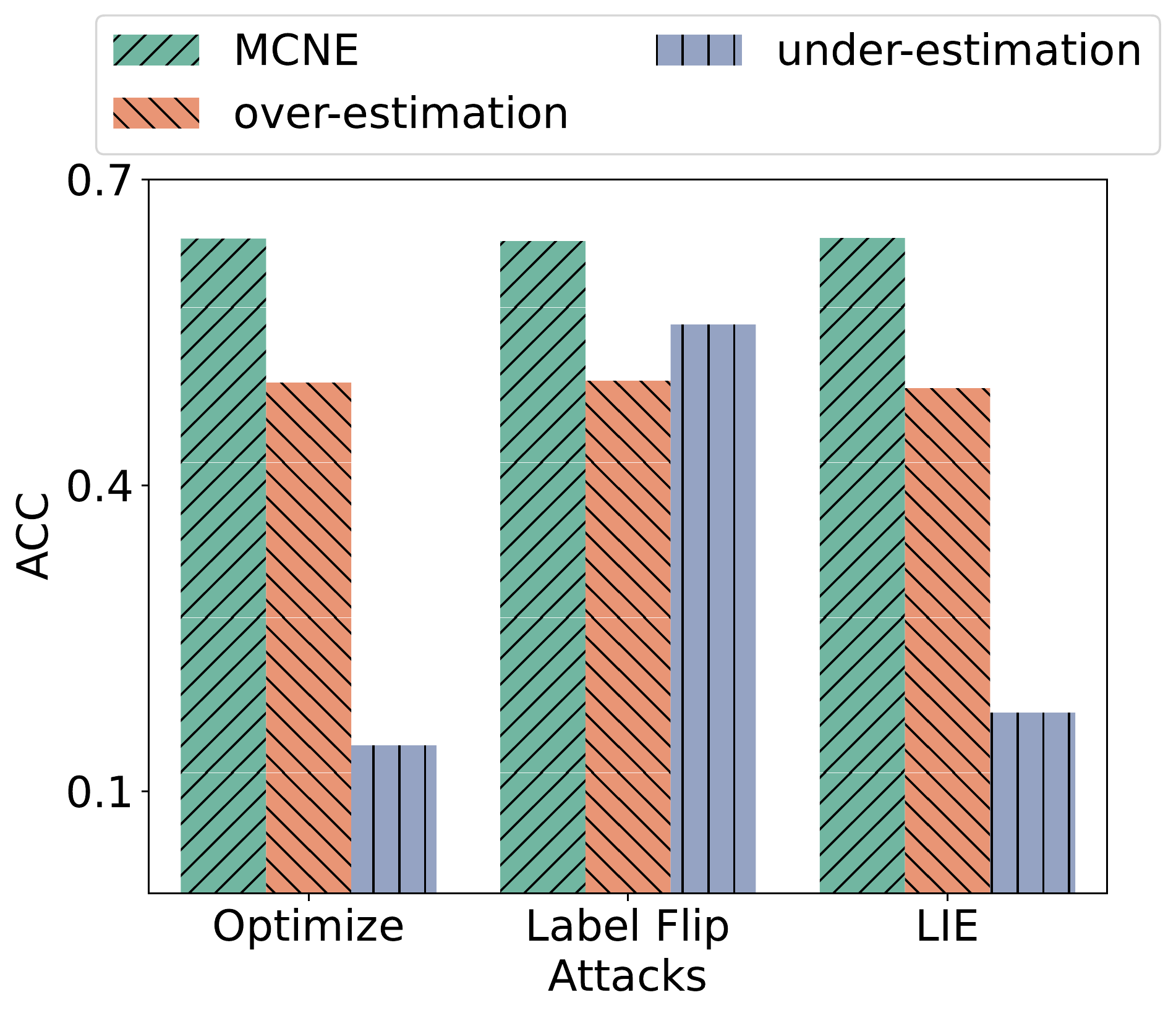}}
     \subfigure[MIND]{\includegraphics[width=0.23\textwidth]{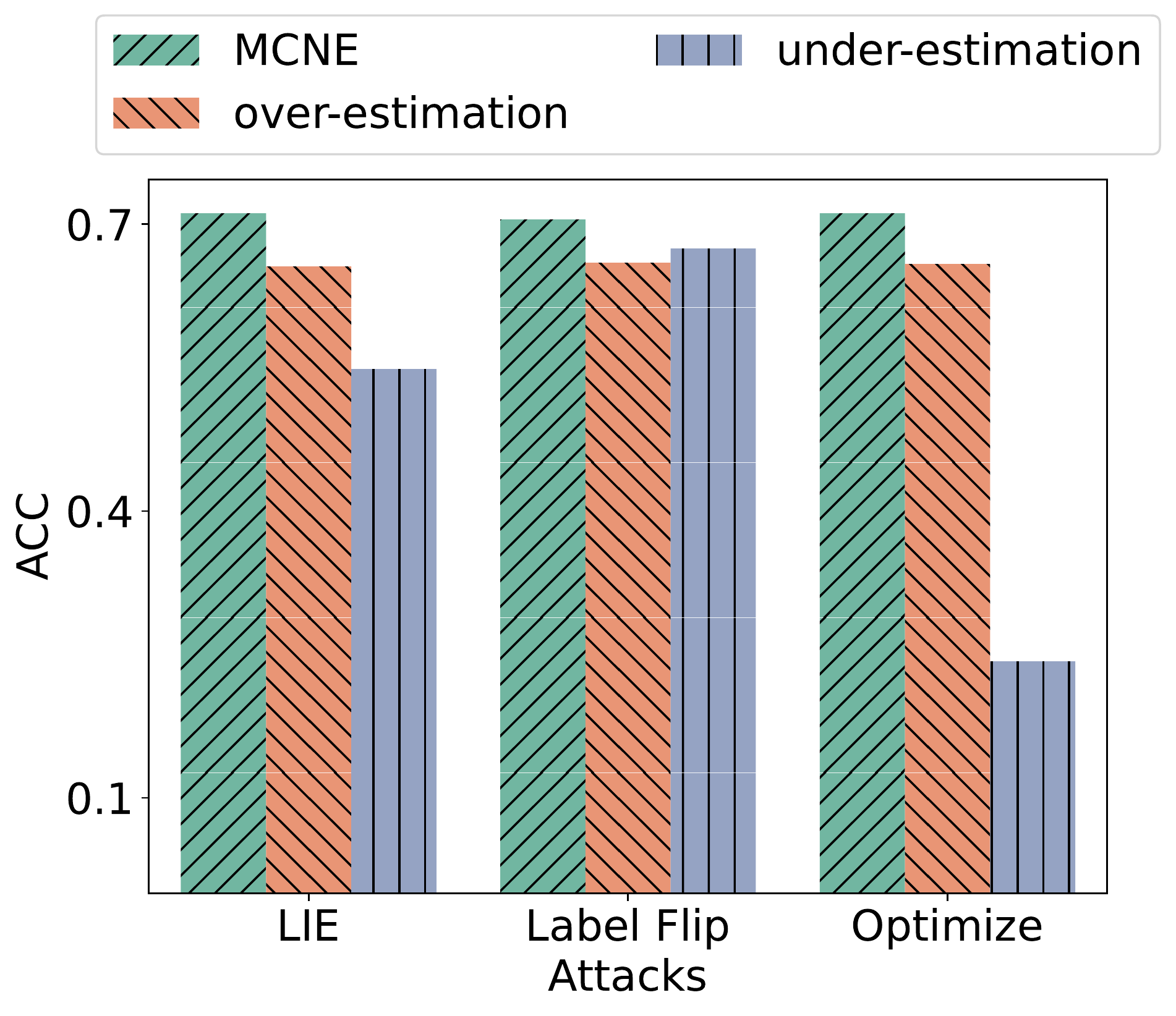}}
     \caption{Results of applying malicious client number estimator, under-estimating and over-estimating the number of malicious clients in the IID dynamic-ratio setting.}
     \label{fig:ablation}
 \end{figure}
 

In this subsection, we compare applying the malicious client number estimator (MCNE) with under-estimating and over-estimating the number of malicious clients $\widetilde{m}$ in the IID dynamic-ratio settings.
For the experiments of under-estimation, we set the estimated number of malicious clients $\widetilde{m}$ as 5, which equals the expectation of selected malicious clients in each round.
For the experiments of over-estimation, we set the estimated number of malicious clients $\widetilde{m}$ as 15, since $p(m > 15) \leqslant 1.54^{-5}$.
Due to the space limitation, we only display the experimental results on CIFAR10 and MIND in Figure~\ref{fig:ablation}.
The results on MNIST are shown in Appendix~\ref{sec:abl-mnist}
The performance of experiments with MCNE is consistently higher than those with over-estimation and under-estimation.
This is because the number of malicious clients follows a hypergeometric distribution in the dynamic-ratio setting, which is hard to be estimated by a fixed parameter.
Under-estimating the number of malicious clients makes some malicious clients selected in some rounds, while over-estimation filters some benign clients in most rounds.

%% file: src/conclusion.tex
\section{Conclusion}
In this paper, we propose FedRA, a robust quantity-aware aggregation method for federated learning.
It aims to aggregate clients' local model updates with awareness of clients' quantities to benefit model performance while being quantity-robust to defend against quantity-enhanced attacks.
FedRA contains two key components, i.e., quantity-robust scorer and malicious client number estimator.
Based on the principle that benign updates with larger quantities should be closer,  the quantity-robust scorer assigns malicious updates higher scores jointly considering received model updates and  quantities.
Since the number of malicious clients varies in different rounds, the malicious client number estimator predicts the number of clients by maximizing the log-likelihood of quantity robust scores.
We theoretically prove that our FedRA is quantity-robust.
Meanwhile, experiments on three public datasets demonstrate the effectiveness of our FedRA.

%% file: src/supp.tex
\setcounter{equation}{0}
\input{src/supp/fedra}

\input{src/supp/exp}

%% file: src/supp/fedra.tex
\section{Therorical Proof}
\subsection{Proof of Proposition 1}
\label{sec:proof-fedra}
\textit{Proof.}
We analyze benign $i_*$ and malicious $i_*$ separately.
\begin{equation}
    \begin{aligned}
    \E[\sqrt{q}||\textbf{g} - \boldsymbol \mu ||_1] &\leq \sum_{\text{benign $i$}} \E[\sqrt{q^i}||\textbf{g}^i -\boldsymbol \mu||_1 \mathbb{I}(i_* = i)] \\
    &+ \sum_{\text{malicious $k$}} \E[\sqrt{q^k}||\textbf{g}^k -\boldsymbol \mu||_1 \mathbb{I}(i_* = k)],
    \end{aligned}
    \label{eq:pro3-error}
\end{equation}
where $q$ and $\textbf{g}$ are the quantity and update of selected client $i_*$.

\begin{lemma}
Let $\textbf{g}^1$, ..., $\textbf{g}^{n}$ be independent identically distributed random vectors with the same quantity $q$, where $\textbf{g}^i \subseteq \mathbb{R}^d$, and $\E\, [\textbf{g}^i] =\boldsymbol \mu$.
Supposing that Assumption~\ref{assume:sample-sub-exp} holds,
then we have
\begin{equation}
\E[\max_{i}||\textbf{g}^i -\boldsymbol \mu||_1]\leq \sqrt{2 \ln 2n} ||\boldsymbol \sigma||_1 / \sqrt{q},
\end{equation}
where $\boldsymbol \sigma$ is a d-dimensional vector denoted as $[\sigma_1, ..., \sigma_{d}]$.
\label{lemma:exp-sub-exp-mu}
\end{lemma}

Lemma~\ref{lemma:exp-sub-exp-mu} is proved in Appendix~\ref{sec:proof-exp-sub-exp-mu}.
When $i_*$ is benign, according to Lemma~\ref{lemma:low-variance} and Lemma~\ref{lemma:exp-sub-exp-mu}, we have
\begin{equation}
    \begin{aligned}
    \sum_{\text{benign $i$}} \E[\sqrt{q^i}||\textbf{g}^i -\boldsymbol \mu||_1 \mathbb{I}(i_* = i)] &\leq \E[\max_{\text{benign $i$}}\sqrt{q^i}||\textbf{g}^i -\boldsymbol \mu||_1] \\
    &\leq \sqrt{2 \ln 2(n-m)} ||\boldsymbol \sigma||_1.
    \end{aligned}
\end{equation}

When $i_*$ is malicious, the error can be formulated as
\begin{equation}
    \begin{aligned}
    &\sum_{\text{malicious $k$}} \E[\sqrt{q^k}||\textbf{g}^k -\boldsymbol \mu||_1 \mathbb{I}(i_* = k)]  \\
    &\leq \underbrace{\sum_{\text{malicious $k$}}\E[\sqrt{q^k}||\textbf{g}^k - \frac{1}{\displaystyle\sum_{j:i_* \to j \atop \text{benign $j$}}q^j}\sum_{j:i_* \to j \atop \text{benign $j$}}q^j\textbf{g}^j||_1\mathbb{I}(i_* = k)]}_{\text{Term 1}} \\
    &+ \underbrace{\sum_{\text{malicious $k$}}\E[\sqrt{q^k}||(\frac{1}{\displaystyle\sum_{j:i_* \to j \atop \text{benign $j$}}q^j}\sum_{j:i_* \to j \atop \text{benign $j$}}q^j\textbf{g}^j) - \boldsymbol \mu||_1\mathbb{I}(i_* = k)]}_{\text{Term 2}}.
    \end{aligned}
\end{equation}

For Term 1, we have
\begin{equation}
\begin{aligned}
&\sum_{\text{malicious $k$}}\E[\sqrt{q^k}||\textbf{g}^k - \frac{1}{\displaystyle\sum_{j:i_* \to j \atop \text{benign $j$}}q^j}\sum_{j:i_* \to j \atop \text{benign $j$}}q^j\textbf{g}^j||_1\mathbb{I}(i_* = k)] \\
&\leq \sum_{\text{malicious $k$}}\E[\frac{\sqrt{q^k}}{\displaystyle\sum_{j:i_* \to j \atop \text{benign $j$}}q^j}\sum_{j:i_* \to j \atop \text{benign $j$}}q^j||\textbf{g}^k - \textbf{g}^j||_1\mathbb{I}(i_* = k)]
\end{aligned}
\end{equation}

Denote $\mathcal{B}$ as the (n-m-2) benign clients with smallest quantities. Since $\sqrt{\frac{q^k}{q^j(q^j +q^k)}} > \frac{1}{q^j + 1}$, we have
\begin{equation}
\begin{aligned}
&\sum_{\text{malicious $k$}}\E[\sqrt{q^k}||\textbf{g}^k - \frac{1}{\displaystyle\sum_{j:i_* \to j \atop \text{benign $j$}}q^j}\sum_{j:i_* \to j \atop \text{benign $j$}}q^j\textbf{g}^j||_1\mathbb{I}(i_* = k)] \\
&\leq \frac{\displaystyle\max_{\text{benign $i$}}[q^i + 1]}{\sum_{i \in \mathcal{B}}q^i}\sum_{\text{malicious $k$}} \\
&\E[\sqrt{q^k}\sum_{j:i_* \to j \atop \text{benign $j$}}\sqrt{\frac{q^k q^j}{q^k+q^j}}||\textbf{g}^k - \textbf{g}^j||_1\mathbb{I}(i_* = k)].
\end{aligned}
\label{eq:pro3-benign-mal}
\end{equation}

If $i_* = k$ is selected, it implies for any correct index $i$
\begin{equation}
\begin{aligned}
    &\sum_{\text{malicious $k$}}\sqrt{q^k}[\sum_{j:i_* \to j \atop \text{benign $j$}}\sqrt{\frac{q^k q^j}{q^k+q^j}}||\textbf{g}^k - \textbf{g}^j||_1 \\ 
    &+ \sum_{j:i_* \to j \atop \text{malicious $j$}}\sqrt{\frac{q^k q^j}{q^k+q^j}}||\textbf{g}^k - \textbf{g}^j||_1]\mathbb{I}(i_* = k) \\
    &\leq \sqrt{q^i}\sum_{j:i \to j \atop \text{benign $j$}}\sqrt{\frac{q^i q^j}{q^i+q^j}}||\textbf{g}^i - \textbf{g}^j||_1 \\
    &+ \sqrt{q^i}\sum_{j:i \to j \atop \text{malicious $j$}}\sqrt{\frac{q^i q^j}{q^i+q^j}}||\textbf{g}^i - \textbf{g}^j||_1.
\label{eq:fedra-mal-q}
\end{aligned}
\end{equation}

We focus on the second term of Equation~\ref{eq:fedra-mal-q}. Since any correct index $i$ has $n-m-2$ neighbors and $m+1$ non-neighbors.
There exists at least one benign index $\varsigma(i)$, which has Q value score than any of its neighbors.
Therefore, $\forall \; j:i \to \text{malicious $j$}$, $\sqrt{\frac{q^i q^j}{q^i+q^j}}||\textbf{g}^i - \textbf{g}^j||_1 \leq \sqrt{\frac{q^i q^{\varsigma(i)}}{q^i+q^{\varsigma(i)}}}||\textbf{g}^i - \textbf{g}^{\varsigma(i)}||_1$.
Then we have

\begin{equation}
    \begin{aligned} 
    &\sqrt{q^i}\sum_{j:i \to j \atop \text{malicious $j$}}\sqrt{\frac{q^i q^j}{q^i+q^j}}||\textbf{g}^i - \textbf{g}^j||_1 \\
    & \leq m \sqrt{q^i} \sqrt{\frac{q^i q^{\varsigma(i)}}{q^i+q^{\varsigma(i)}}}||\textbf{g}^i - \textbf{g}^{\varsigma(i)}||_1.
    \end{aligned}
\label{eq:fedra-farthest-neighbor}
\end{equation}

With Lemma~\ref{lemma:sub-exp-quantity}, bringing Equation~\ref{eq:fedra-farthest-neighbor} and~\ref{eq:fedra-mal-q} back to Equation~\ref{eq:pro3-benign-mal}, we can obtain

\begin{equation}
    \begin{aligned}
    &\sum_{\text{malicious $k$}}\E[\sqrt{q^k}||\textbf{g}^k - \overline{\textbf{g}}||_1\mathbb{I}(i_* = k)] \leqslant A_{n,m} B_q ||\boldsymbol \sigma||_1,
    \end{aligned}
\end{equation}
where 
\begin{equation}
    \begin{aligned}
    A_{n,m} &= \sqrt{2\ln 2}(n-m) + m\sqrt{2 \ln 2(n-m)}, \\
    B_q &= \frac{\max_{\text{benign $i$}}[q^i + 1] \min_{\text{benign $i$}}}{\sum_{i \in \mathcal{B}}q^i}, \\
    \overline{\textbf{g}} &= \frac{1}{\displaystyle\sum_{j:i_* \to j \atop \text{benign $j$}}q^j}\sum_{j:i_* \to j \atop \text{benign $j$}}q^j\textbf{g}^j. 
    \end{aligned}
\end{equation}

It is easy to prove the following conclusion with the same logic.
\begin{equation}
    \begin{aligned}
    &\E[\sqrt{q}||\textbf{g} - \overline{\textbf{g}}||_1 \leqslant A_{n,m} B_q ||\boldsymbol \sigma||_1.
    \end{aligned}
\end{equation}

For Term 2, according to Lemma~\ref{lemma:low-variance} we have 
\begin{equation}
    \begin{aligned}
        &\sum_{\text{malicious $k$}}\E[\sqrt{q^k}||\overline{\textbf{g}} - \boldsymbol \mu||_1\mathbb{I}(i_* = k)] \\
        &\leq \sqrt{\frac{2 \ln 2 }{\sum_{j:i_* \to j \atop \text{benign $j$}}q^j}} \max_{\text{malicious $i$}}[\sqrt{q^i}]||\boldsymbol \sigma||_1 \\
        &\leq \sqrt{\frac{2 \ln 2 }{\sum_{i \in \mathcal{B}}q^i}} \max_{\text{malicious $i$}}[\sqrt{q^i}]||\boldsymbol \sigma||_1.
    \end{aligned}
\end{equation}
Therefore, if all malicious quantities $q^i$ satisfy $q^i \leq \sum_{i \in \mathcal{B}}q^i$, then we have 
\begin{equation}
    \begin{aligned}
        \sum_{\text{malicious $k$}}\E[\sqrt{q^k}||\overline{\textbf{g}} - \boldsymbol \mu||_1\mathbb{I}(i_* = k)] &\leq \sqrt{2 \ln 2 }||\boldsymbol \sigma||_1.
    \end{aligned}
\end{equation}

Bring everything back to Equation~\ref{eq:pro3-error}, we have
\begin{equation}
    \begin{aligned}
        \E[\sqrt{q}||\textbf{g} - \boldsymbol \mu||_1] \leqslant  (A_{n,m}B_q + C_{n,m})||\boldsymbol \sigma||_1,
    \end{aligned}
\end{equation}
where $C_{n,m} = \sqrt{2 \ln 2(n-m)} + \sqrt{2 \ln 2}$.

\subsection{Proof of Theorem 2}
\label{sec:proof-non-convex-fedra}
\textit{Proof.}
Using the smoothness of $F(\cdot)$, we have
\begin{equation}
    \begin{aligned}
    F(\textbf{w}_{t+1}) &\leq F(\textbf{w}_t) + \langle\nabla F(\textbf{w}_t), \textbf{w}_{t+1} - \textbf{w}_{t}\rangle + \frac{L_F}{2}||\textbf{w}_{t+1} - \textbf{w}_t||_2^2 \\
    &= F(\textbf{w}_t) + \frac{\eta^2L_F}{2q_t}||\sqrt{q_t}(\textbf{g}_t- \nabla F(\textbf{w}_t)) + \sqrt{q_t}\nabla F(\textbf{w}_t)||_2^2 \\
    &- \frac{\eta}{q_t}\langle\sqrt{q_t}\nabla F(\textbf{w}_t), \sqrt{q_t}(\textbf{g}_t - \nabla F(\textbf{w}_t)) + \sqrt{q_t}\nabla F(\textbf{w}_t) \rangle \\
    &= F(\textbf{w}_t) + (\frac{\eta^2L_F}{2} - \eta)||\nabla F(\textbf{w}_t)||_2^2 \\
    &+ \frac{\eta^2L_F}{2q_t}||\sqrt{q_t}(\textbf{g}_t- \nabla F(\textbf{w}_t))||_2^2 \\
    & + (\frac{\eta^2L_F-\eta}{q_t})\langle\sqrt{q_t}\nabla F(\textbf{w}_t), \sqrt{q_t}(\textbf{g}_t - \nabla F(\textbf{w}_t)) \rangle.
    \end{aligned}
\end{equation}

Let $\eta = \frac{1}{L_F}$. According to Proposition~\ref{prop:fedra}, we further obtain
\begin{equation}
    \begin{aligned}
    &\E [F(\textbf{w}_{t+1}) - F(\textbf{w}^*)] \leq \E [F(\textbf{w}_{t}) - F(\textbf{w}^*)] \\
    &- \frac{1}{2L_F}\E[||\nabla F(\textbf{w}_t)||_2^2] + \frac{1}{2q_tL_F}((A_{n,m}B_q + C_{n,m})||\boldsymbol \sigma||_1)^2. 
    \label{eq:smoothness-con-convex-fedra}
    \end{aligned}
\end{equation}
Sum up Equation~\ref{eq:smoothness-con-convex-fedra}, we have 
\begin{equation}
    \begin{aligned}
         0 &\leq \E [F(\textbf{w}_{T}) - F(\textbf{w}^*)] \\ &\leq \E [F(\textbf{w}^{0}) - F(\textbf{w}^*)] - \frac{1}{2L_F}\sum_{t=0}^{T-1}\E[||\nabla F(\textbf{w}_t)||_2^2] \\
         &+ \frac{1}{2 L_F}\sum_{t=0}^{T-1}{\frac{1}{q_t}}((A_{n,m}B_q + C_{n,m})||\boldsymbol \sigma||_1)^2,
    \end{aligned}
\end{equation}
which implies
\begin{equation}
\begin{aligned}
    &\min_{t=0,...,T}\E[||\nabla F(\textbf{w}_t)||_2^2] \leq \frac{2L_F}{T}\E [F(\textbf{w}_{0}) - F(\textbf{w}^*)] \\
    &+ \frac{1}{T}\sum_{t=0}^{T-1}{\frac{1}{q_t}}((A_{n,m}B_q + C_{n,m})||\boldsymbol \sigma||_1)^2.
\end{aligned}
\end{equation}

\subsection{Proof of Lemma 1}
\label{sec:proof-lemma-sub-exp-quantity}
\textit{Proof.}
We first convert the problem into computing expectations of each dimension.
\begin{equation}
\begin{aligned}
\E[||\textbf{g}^i -\textbf{g}^j||_1] &= \E[\sum_{k\in\{1,...,d\}}|g_k^i - g_k^j|] \\
&= \sum_{k\in\{1,...,d\}}\E[|g_k^i - g_k^j|].
\end{aligned} 
\end{equation}

Since any $z \in \mathcal{D}_i$ are independent, if $|\lambda| < \frac{q^i}{\alpha_k}$, we then have
\begin{equation}
    \begin{aligned}
    \E[e^{\lambda(g_k^i - \mu_k)}] &= \E[e^{\frac{\lambda}{q^i}\sum_{z\in \mathcal{D}_i}(\partial_kf(\textbf{w}; z) - \mu_k)}]
    &\leq e^{\frac{\lambda^2\sigma_k^2}{2q^i}},
    \end{aligned} 
\end{equation}
which implies $g_k^i$ is sub-exponential with parameters $(\frac{\sigma_k^2}{q^i}, \frac{\alpha_k}{q^i})$, where $\alpha_k<\frac{\sigma_k}{\sqrt{2 \ln 2n}}$.

Let $|\lambda| < \min(\frac{q^i}{\alpha_k}, \frac{q^j}{\alpha_k})$. Since $g_k^i$ and $g_k^j$ are independently distributed, we can obtain
\begin{equation}
    \begin{aligned}
    \E[e^{\lambda(g_k^i - g_k^j)}] &= \E[e^{\lambda((g_k^i -\mu_k) - (g_k^j - \mu_k))}] \\
    &= \E[e^{\lambda(g_k^i -\mu_k)}]\E[e^{-\lambda(g_k^j -\mu_k)}] \\
    &\leq e^{\frac{\lambda^2\sigma_k^2}{2} \frac{q^i+q^j}{q^iq^j}}.
    \end{aligned}
\end{equation}
Following the same logic in Lemma~\ref{lemma:exp-sub-exp-mu} and setting $\lambda = \frac{\sqrt{2 \ln 2n}}{\sigma_k}\sqrt{\frac{q^iq^j}{q^i+q^j}}$, we can have
\begin{equation}
\E[||\textbf{g}^i - \textbf{g}^j||_1] \leq \sqrt{2 \ln 2}||\boldsymbol \sigma||_1 \sqrt{\frac{q^i+q^j}{q^iq^j}}.
\end{equation}

\subsection{Proof of Lemma 2}
\label{sec:proof-low-variance}
\textit{Proof.}
As defined in Section~\ref{sec:prob-define}, we have $g_k^i = \frac{1}{q^i} \sum_{z \in \mathcal{D}_i} \partial_k f(\textbf{w};z)$.
Therefore, we can obtain
\begin{equation}
    \begin{aligned}
    \E[g_k^i] = \frac{1}{q^i} \sum_{z \in \mathcal{D}_i} \E[\partial_kf(\textbf{w};z)] = \mu_k,
    \end{aligned}
\end{equation}
\begin{equation}
    \begin{aligned}
    \E[(g_k^i - \mu_k))^2] &= \E[(\frac{1}{q^i} \sum_{z \in \mathcal{D}_i} \partial_kf(\textbf{w};z) - \mu_k)^2] \\
    &= \frac{1}{(q^i)^2} \sum_{z \in \mathcal{D}_i} \E[(\partial_kf(\textbf{w};z) - \mu_k)^2] = \frac{\sigma_k^2}{q^i}.
    \end{aligned}
\end{equation}

Since $\partial_k f(\textbf{w};z)$ is sub-exponential with parameters $(v_k^2, \alpha_k)$, we have
\begin{equation}
    \begin{aligned}
    e^{\frac{\lambda}{q^i}\sum_{z \in \mathcal{D}_i} (\partial_k f(\textbf{w};z) - \mu_k)} \leq e^{\frac{1}{2q^i}v^2\lambda^2},
    \end{aligned}
\end{equation}
when $\lambda \leq \frac{q^i}{\alpha_k}$.
It shows $g_k^i$ is sub-exponential with parameters $(\frac{v_k^2}{q^i}, \frac{\alpha_k}{q^i})$, where $v_k = \sigma_k$ and $\alpha_k \leq \frac{\sigma_k}{\sqrt{2\ln 2n}}$. $n$ is the number of clients sampled by server in each round.

Following the same logic in Lemma~\ref{lemma:exp-sub-exp-mu}, we have
\begin{equation}
    \begin{aligned}
    \E[|g_k^i - \mu_k|] &\leq \sqrt{2\ln 2} \frac{\sigma_k}{\sqrt{q^i}}.
    \end{aligned}
\end{equation}

Obviously, we can obtain
\begin{equation}
    \begin{aligned}
        \E[\textbf{g}^i] = \boldsymbol \mu, \;
        \E[\sqrt{q^i}||\textbf{g}^i - \boldsymbol \mu||_1] \leq \sqrt{2\ln 2}||\boldsymbol \sigma||_1.
    \end{aligned}
\end{equation}

\subsection{Proof of Lemma 3}
\label{sec:proof-exp-sub-exp-mu}

\textit{Proof.}
We first convert the problem of computing the expectation of the maximum of the $\normlone$ norm of the d-dimensional vectors into the problem of computing expectations of the maximum of each dimension of the d-dimensional vectors.
\begin{equation}
\begin{aligned}
\label{eq:lemma2-max}
\E[\max_{i}||\textbf{g}^i -\boldsymbol \mu||_1] &= \E[\max_{i}\sum_{k\in\{1,...,d\}}|g_k^i - \mu_k|] \\
&\leq \sum_{k\in\{1,...,d\}}\E[\max_{i}|g_k^i - \mu_k|].
\end{aligned} 
\end{equation}

Following the same logic in Appendix~\ref{sec:proof-low-variance}, it is easy to prove $g_k^i$ is sub-exponential with parameters $(\frac{v_k^2}{q}, \frac{\alpha_k}{q})$, where $v_k = \sigma_k$ and $\alpha_k \leq \frac{\sigma_k}{\sqrt{2\ln 2n}}$.
Define $x_k^i = g_k^i - \mu_k$.
Denote a list of values $X=\{x_k^1, -x_k^1, ..., x_k^n, -x_k^n\}$, and $z_k = \max_{i}|g_k^i - \mu_k| = \max_{x \in X}x$.
We then obtain
\begin{equation}
\begin{aligned}
e^{\lambda\E[z_k]} &\leq \E [e^{\lambda z_k}] = \E [\max_{x\in X}e^{\lambda x}] \leq \sum_{x \in X} \E [e^{\lambda x}] \leq 2n e^{\frac{\lambda^2 \sigma_k^2}{2q}}.
\end{aligned} 
\end{equation}
\begin{equation}
\E[z_k] \leq \frac{\ln 2n}{\lambda} + \frac{\lambda \sigma_k^2}{2q}.
\end{equation}

Setting $\lambda = \frac{\sqrt{2q \ln 2n}}{\sigma_k} < \frac{q}{\alpha_k}$, we can get
\begin{equation}
~\label{eq:lemma2-lnn}
\E[z_k] \leq \sigma_k \sqrt{2 \ln 2n} / \sqrt{q}.
\end{equation}

Putting Equation~\ref{eq:lemma2-lnn} back to Equation~\ref{eq:lemma2-max}, we obtain 
\begin{equation}
\E [\max_{i}||\textbf{g}^i -\boldsymbol \mu||_1]\leq \sqrt{2 \ln 2n} ||\boldsymbol \sigma||_1 / \sqrt{q},
\end{equation}
where $\boldsymbol \sigma$ is a d-dimensional vector denoted as $[\sigma_1, ..., \sigma_{d}]$.

%% file: src/supp/exp.tex
\section{Additional Experiment}

\subsection{Distributions of Client Scores}
\label{sec:evidence-mcne}
\begin{figure*}[!t]
  \centering
  \includegraphics[width=0.80\textwidth]{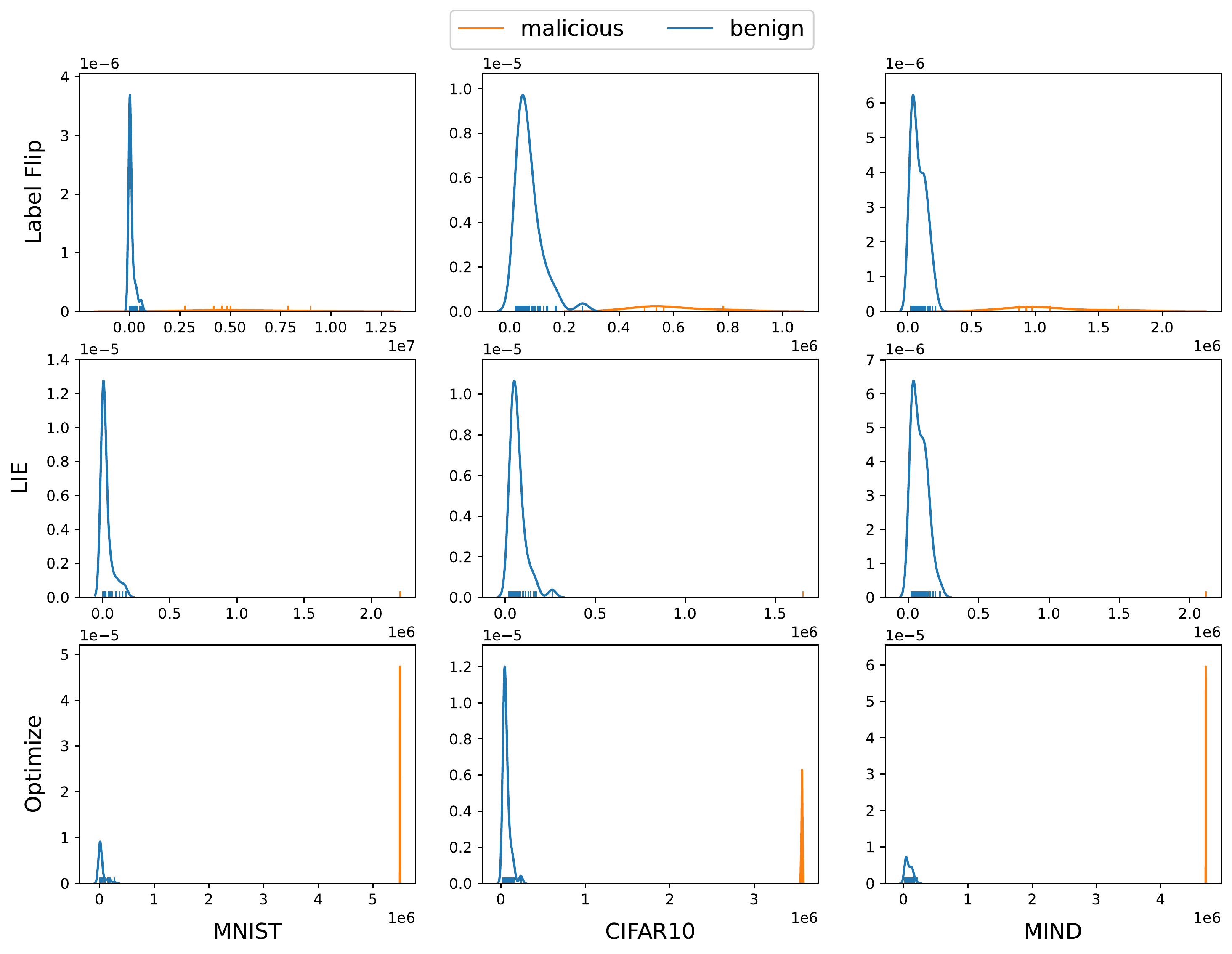}
  \caption{Distributions of client scores in IID dynamic-ratio settings at 1,000 steps.}
  \label{fig:evidence-mcne}
\end{figure*}

In this subsection, we show some distributions of client scores in our experiments.
The distributions support the assumptions in our malicious client number estimator that the scores of benign and malicious clients follow two independent Gaussian distributions, and malicious clients get the largest scores.
In Figure~\ref{fig:evidence-mcne}, we show the distributions of client scores with quantity-enlarge factor $\alpha_q=10$ at 1,000 steps for MIND, MNIST and CIFAR10 dataset respectively.


\section{Ablation Study on MNIST}
\label{sec:abl-mnist}
\begin{figure}[!t]
  \centering
  \includegraphics[width=0.25\textwidth]{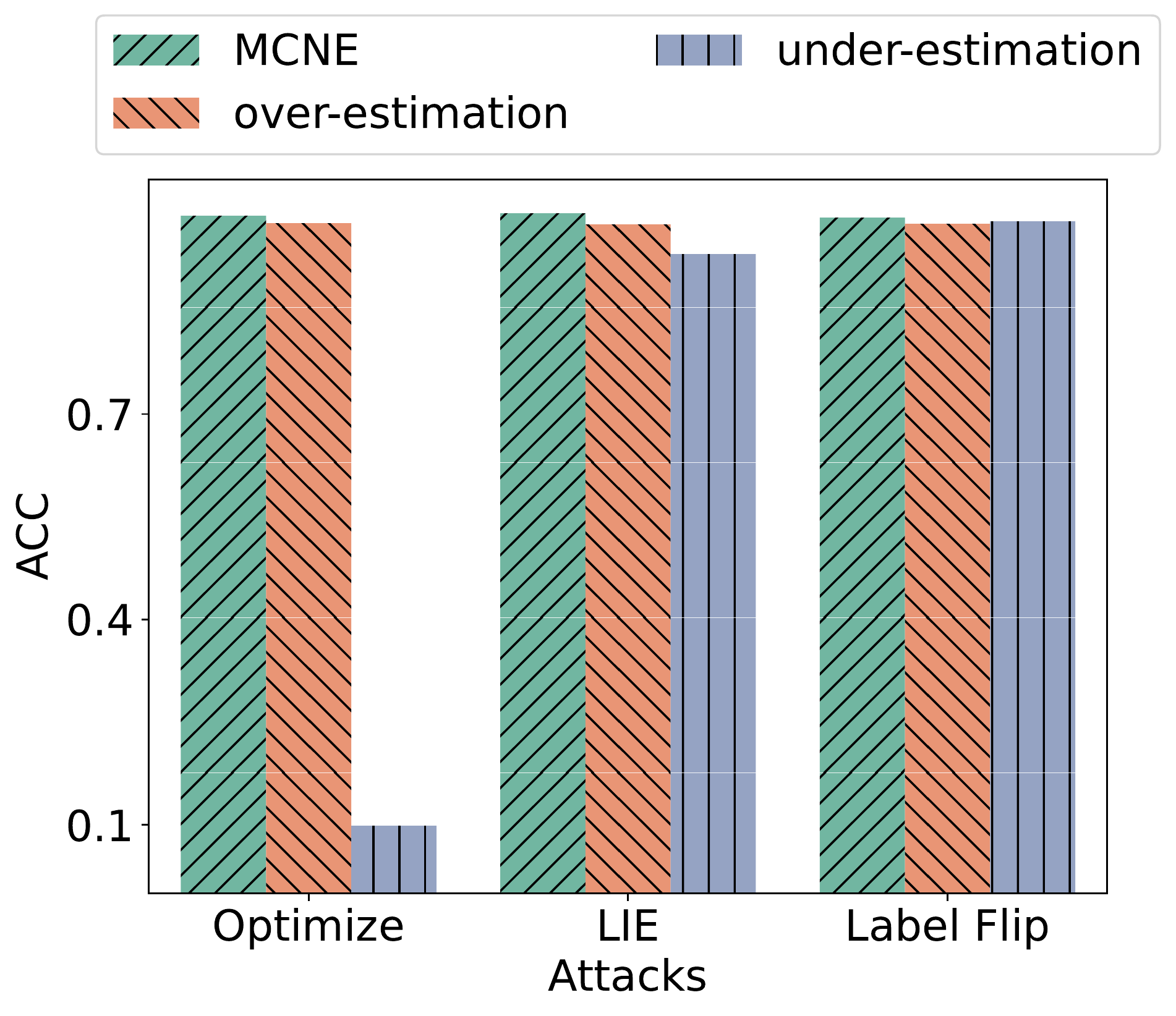}
  \caption{Results of applying malicious client number estimator, under-estimating and over-estimating the number of malicious clients in the IID dynamic-ratio setting on MNIST.}
  \label{fig:abl-mnist}
\end{figure}

In this subsection, we compare applying the malicious client number estimator (MCNE) to under-estimating and over-estimating the number of malicious clients $\widetilde{m}$ in the IID dynamic-ratio settings on MNIST.
The results are shown in Figure~\ref{fig:abl-mnist}
The performance of experiments with MCNE is consistently higher than those with over-estimation and under-estimation.
This is because the number of malicious clients follows a hypergeometric distribution in the dynamic-ratio setting, which is hard to be estimated by a fixed parameter.
Under-estimating the number of malicious clients makes some malicious clients selected in some rounds, while over-estimation filters some benign clients in most rounds.

\section{Experimental Environments}
\label{sec:env}
We conduct experiments on a single V100 GPU with 32GB memory.
The version of CUDA is 11.1.
We use pytorch 1.9.1.

%% file: sample-sigconf.bbl

\begin{thebibliography}{43}


\ifx \showCODEN    \undefined \def \showCODEN     #1{\unskip}     \fi
\ifx \showDOI      \undefined \def \showDOI       #1{#1}\fi
\ifx \showISBNx    \undefined \def \showISBNx     #1{\unskip}     \fi
\ifx \showISBNxiii \undefined \def \showISBNxiii  #1{\unskip}     \fi
\ifx \showISSN     \undefined \def \showISSN      #1{\unskip}     \fi
\ifx \showLCCN     \undefined \def \showLCCN      #1{\unskip}     \fi
\ifx \shownote     \undefined \def \shownote      #1{#1}          \fi
\ifx \showarticletitle \undefined \def \showarticletitle #1{#1}   \fi
\ifx \showURL      \undefined \def \showURL       {\relax}        \fi
\providecommand\bibfield[2]{#2}
\providecommand\bibinfo[2]{#2}
\providecommand\natexlab[1]{#1}
\providecommand\showeprint[2][]{arXiv:#2}

\bibitem[Bagdasaryan et~al\mbox{.}(2020)]%
        {bagdasaryan2020backdoor}
\bibfield{author}{\bibinfo{person}{Eugene Bagdasaryan},
  \bibinfo{person}{Andreas Veit}, \bibinfo{person}{Yiqing Hua},
  \bibinfo{person}{Deborah Estrin}, {and} \bibinfo{person}{Vitaly Shmatikov}.}
  \bibinfo{year}{2020}\natexlab{}.
\newblock \showarticletitle{How to backdoor federated learning}. In
  \bibinfo{booktitle}{\emph{AISTATS}}. \bibinfo{pages}{2938--2948}.
\newblock


\bibitem[Baruch et~al\mbox{.}(2019)]%
        {NEURIPS2019_ec1c5914}
\bibfield{author}{\bibinfo{person}{Gilad Baruch}, \bibinfo{person}{Moran
  Baruch}, {and} \bibinfo{person}{Yoav Goldberg}.}
  \bibinfo{year}{2019}\natexlab{}.
\newblock \showarticletitle{A Little Is Enough: Circumventing Defenses For
  Distributed Learning}. In \bibinfo{booktitle}{\emph{NIPS}},
  Vol.~\bibinfo{volume}{32}.
\newblock


\bibitem[Bhagoji et~al\mbox{.}(2019)]%
        {bhagoji2019analyzing}
\bibfield{author}{\bibinfo{person}{Arjun~Nitin Bhagoji},
  \bibinfo{person}{Supriyo Chakraborty}, \bibinfo{person}{Prateek Mittal},
  {and} \bibinfo{person}{Seraphin Calo}.} \bibinfo{year}{2019}\natexlab{}.
\newblock \showarticletitle{Analyzing federated learning through an adversarial
  lens}. In \bibinfo{booktitle}{\emph{ICML}}. \bibinfo{pages}{634--643}.
\newblock


\bibitem[Blanchard et~al\mbox{.}(2017)]%
        {blanchard2017machine}
\bibfield{author}{\bibinfo{person}{Peva Blanchard}, \bibinfo{person}{El~Mahdi
  El~Mhamdi}, \bibinfo{person}{Rachid Guerraoui}, {and} \bibinfo{person}{Julien
  Stainer}.} \bibinfo{year}{2017}\natexlab{}.
\newblock \showarticletitle{Machine learning with adversaries: Byzantine
  tolerant gradient descent}.
\newblock \bibinfo{journal}{\emph{NIPS}}  \bibinfo{volume}{30}
  (\bibinfo{year}{2017}).
\newblock


\bibitem[Chen et~al\mbox{.}(2019)]%
        {chen2019federated}
\bibfield{author}{\bibinfo{person}{Mingqing Chen}, \bibinfo{person}{Rajiv
  Mathews}, \bibinfo{person}{Tom Ouyang}, {and}
  \bibinfo{person}{Fran{\c{c}}oise Beaufays}.} \bibinfo{year}{2019}\natexlab{}.
\newblock \showarticletitle{Federated learning of out-of-vocabulary words}.
\newblock \bibinfo{journal}{\emph{arXiv preprint arXiv:1903.10635}}
  (\bibinfo{year}{2019}).
\newblock


\bibitem[Chen et~al\mbox{.}(2020)]%
        {chen2020esam}
\bibfield{author}{\bibinfo{person}{Zhihong Chen}, \bibinfo{person}{Rong Xiao},
  \bibinfo{person}{Chenliang Li}, \bibinfo{person}{Gangfeng Ye},
  \bibinfo{person}{Haochuan Sun}, {and} \bibinfo{person}{Hongbo Deng}.}
  \bibinfo{year}{2020}\natexlab{}.
\newblock \showarticletitle{Esam: Discriminative domain adaptation with
  non-displayed items to improve long-tail performance}. In
  \bibinfo{booktitle}{\emph{SIGIR}}. \bibinfo{pages}{579--588}.
\newblock


\bibitem[El~Mhamdi et~al\mbox{.}(2018)]%
        {pmlr-v80-mhamdi18a}
\bibfield{author}{\bibinfo{person}{El~Mahdi El~Mhamdi}, \bibinfo{person}{Rachid
  Guerraoui}, {and} \bibinfo{person}{S{\'e}bastien Rouault}.}
  \bibinfo{year}{2018}\natexlab{}.
\newblock \showarticletitle{The Hidden Vulnerability of Distributed Learning in
  {B}yzantium}. In \bibinfo{booktitle}{\emph{ICML}}, Vol.~\bibinfo{volume}{80}.
  \bibinfo{pages}{3521--3530}.
\newblock


\bibitem[Fang et~al\mbox{.}(2020)]%
        {10.5555/3489212.3489304}
\bibfield{author}{\bibinfo{person}{Minghong Fang}, \bibinfo{person}{Xiaoyu
  Cao}, \bibinfo{person}{Jinyuan Jia}, {and} \bibinfo{person}{Neil~Zhenqiang
  Gong}.} \bibinfo{year}{2020}\natexlab{}.
\newblock \showarticletitle{Local Model Poisoning Attacks to Byzantine-Robust
  Federated Learning}. In \bibinfo{booktitle}{\emph{USENIX}}.
\newblock


\bibitem[Hard et~al\mbox{.}(2018)]%
        {hard2018federated}
\bibfield{author}{\bibinfo{person}{Andrew Hard}, \bibinfo{person}{Kanishka
  Rao}, \bibinfo{person}{Rajiv Mathews}, \bibinfo{person}{Swaroop Ramaswamy},
  \bibinfo{person}{Fran{\c{c}}oise Beaufays}, \bibinfo{person}{Sean
  Augenstein}, \bibinfo{person}{Hubert Eichner}, \bibinfo{person}{Chlo{\'e}
  Kiddon}, {and} \bibinfo{person}{Daniel Ramage}.}
  \bibinfo{year}{2018}\natexlab{}.
\newblock \showarticletitle{Federated learning for mobile keyboard prediction}.
\newblock \bibinfo{journal}{\emph{arXiv preprint arXiv:1811.03604}}
  (\bibinfo{year}{2018}).
\newblock


\bibitem[Karimireddy et~al\mbox{.}(2020)]%
        {karimireddy2020scaffold}
\bibfield{author}{\bibinfo{person}{Sai~Praneeth Karimireddy},
  \bibinfo{person}{Satyen Kale}, \bibinfo{person}{Mehryar Mohri},
  \bibinfo{person}{Sashank Reddi}, \bibinfo{person}{Sebastian Stich}, {and}
  \bibinfo{person}{Ananda~Theertha Suresh}.} \bibinfo{year}{2020}\natexlab{}.
\newblock \showarticletitle{Scaffold: Stochastic controlled averaging for
  federated learning}. In \bibinfo{booktitle}{\emph{ICML}}.
  \bibinfo{pages}{5132--5143}.
\newblock


\bibitem[Kim et~al\mbox{.}(2019)]%
        {kim2019sequential}
\bibfield{author}{\bibinfo{person}{Yejin Kim}, \bibinfo{person}{Kwangseob Kim},
  \bibinfo{person}{Chanyoung Park}, {and} \bibinfo{person}{Hwanjo Yu}.}
  \bibinfo{year}{2019}\natexlab{}.
\newblock \showarticletitle{Sequential and Diverse Recommendation with Long
  Tail.}. In \bibinfo{booktitle}{\emph{IJCAI}}, Vol.~\bibinfo{volume}{19}.
  \bibinfo{pages}{2740--2746}.
\newblock


\bibitem[Krizhevsky et~al\mbox{.}(2009)]%
        {krizhevsky2009learning}
\bibfield{author}{\bibinfo{person}{Alex Krizhevsky}, \bibinfo{person}{Geoffrey
  Hinton}, {et~al\mbox{.}}} \bibinfo{year}{2009}\natexlab{}.
\newblock \showarticletitle{Learning multiple layers of features from tiny
  images}.
\newblock  (\bibinfo{year}{2009}).
\newblock


\bibitem[Kumar and Singla(2021)]%
        {kumar2021federated}
\bibfield{author}{\bibinfo{person}{Yogesh Kumar} {and} \bibinfo{person}{Ruchi
  Singla}.} \bibinfo{year}{2021}\natexlab{}.
\newblock \showarticletitle{Federated learning systems for healthcare:
  perspective and recent progress}.
\newblock \bibinfo{journal}{\emph{Federated Learning Systems: Towards
  Next-Generation AI}} (\bibinfo{year}{2021}), \bibinfo{pages}{141--156}.
\newblock


\bibitem[LeCun et~al\mbox{.}(1998)]%
        {lecun1998gradient}
\bibfield{author}{\bibinfo{person}{Yann LeCun}, \bibinfo{person}{L{\'e}on
  Bottou}, \bibinfo{person}{Yoshua Bengio}, {and} \bibinfo{person}{Patrick
  Haffner}.} \bibinfo{year}{1998}\natexlab{}.
\newblock \showarticletitle{Gradient-based learning applied to document
  recognition}.
\newblock \bibinfo{journal}{\emph{Proc. IEEE}} \bibinfo{volume}{86},
  \bibinfo{number}{11} (\bibinfo{year}{1998}), \bibinfo{pages}{2278--2324}.
\newblock


\bibitem[Li et~al\mbox{.}(2017)]%
        {10.1145/3123266.3123316}
\bibfield{author}{\bibinfo{person}{Jingjing Li}, \bibinfo{person}{Ke Lu},
  \bibinfo{person}{Zi Huang}, {and} \bibinfo{person}{Heng~Tao Shen}.}
  \bibinfo{year}{2017}\natexlab{}.
\newblock \showarticletitle{Two Birds One Stone: On Both Cold-Start and
  Long-Tail Recommendation}. In \bibinfo{booktitle}{\emph{MM}}.
  \bibinfo{pages}{898–906}.
\newblock


\bibitem[Li et~al\mbox{.}(2019)]%
        {li2019both}
\bibfield{author}{\bibinfo{person}{Jingjing Li}, \bibinfo{person}{Ke Lu},
  \bibinfo{person}{Zi Huang}, {and} \bibinfo{person}{Heng~Tao Shen}.}
  \bibinfo{year}{2019}\natexlab{}.
\newblock \showarticletitle{On both cold-start and long-tail recommendation
  with social data}.
\newblock \bibinfo{journal}{\emph{TKDE}} \bibinfo{volume}{33},
  \bibinfo{number}{1} (\bibinfo{year}{2019}), \bibinfo{pages}{194--208}.
\newblock


\bibitem[Li et~al\mbox{.}(2020a)]%
        {li2020review}
\bibfield{author}{\bibinfo{person}{Li Li}, \bibinfo{person}{Yuxi Fan},
  \bibinfo{person}{Mike Tse}, {and} \bibinfo{person}{Kuo-Yi Lin}.}
  \bibinfo{year}{2020}\natexlab{a}.
\newblock \showarticletitle{A review of applications in federated learning}.
\newblock \bibinfo{journal}{\emph{Computers \& Industrial Engineering}}
  \bibinfo{volume}{149} (\bibinfo{year}{2020}), \bibinfo{pages}{106854}.
\newblock


\bibitem[Li et~al\mbox{.}(2020b)]%
        {li2020federated}
\bibfield{author}{\bibinfo{person}{Tian Li}, \bibinfo{person}{Anit~Kumar Sahu},
  \bibinfo{person}{Manzil Zaheer}, \bibinfo{person}{Maziar Sanjabi},
  \bibinfo{person}{Ameet Talwalkar}, {and} \bibinfo{person}{Virginia Smith}.}
  \bibinfo{year}{2020}\natexlab{b}.
\newblock \showarticletitle{Federated optimization in heterogeneous networks}.
\newblock \bibinfo{journal}{\emph{Proceedings of Machine Learning and Systems}}
   \bibinfo{volume}{2} (\bibinfo{year}{2020}), \bibinfo{pages}{429--450}.
\newblock


\bibitem[Liu et~al\mbox{.}(2017)]%
        {liu2017trojaning}
\bibfield{author}{\bibinfo{person}{Yingqi Liu}, \bibinfo{person}{Shiqing Ma},
  \bibinfo{person}{Yousra Aafer}, \bibinfo{person}{Wen-Chuan Lee},
  \bibinfo{person}{Juan Zhai}, \bibinfo{person}{Weihang Wang}, {and}
  \bibinfo{person}{Xiangyu Zhang}.} \bibinfo{year}{2017}\natexlab{}.
\newblock \showarticletitle{Trojaning attack on neural networks}. In
  \bibinfo{booktitle}{\emph{NDSS}}.
\newblock


\bibitem[Long et~al\mbox{.}(2022)]%
        {long2022retrieval}
\bibfield{author}{\bibinfo{person}{Alexander Long}, \bibinfo{person}{Wei Yin},
  \bibinfo{person}{Thalaiyasingam Ajanthan}, \bibinfo{person}{Vu Nguyen},
  \bibinfo{person}{Pulak Purkait}, \bibinfo{person}{Ravi Garg},
  \bibinfo{person}{Alan Blair}, \bibinfo{person}{Chunhua Shen}, {and}
  \bibinfo{person}{Anton van~den Hengel}.} \bibinfo{year}{2022}\natexlab{}.
\newblock \showarticletitle{Retrieval augmented classification for long-tail
  visual recognition}. In \bibinfo{booktitle}{\emph{CVPR}}.
  \bibinfo{pages}{6959--6969}.
\newblock


\bibitem[McMahan et~al\mbox{.}(2017)]%
        {mcmahan2017communication}
\bibfield{author}{\bibinfo{person}{Brendan McMahan}, \bibinfo{person}{Eider
  Moore}, \bibinfo{person}{Daniel Ramage}, \bibinfo{person}{Seth Hampson},
  {and} \bibinfo{person}{Blaise~Aguera y Arcas}.}
  \bibinfo{year}{2017}\natexlab{}.
\newblock \showarticletitle{Communication-efficient learning of deep networks
  from decentralized data}. In \bibinfo{booktitle}{\emph{AISTATS}}.
  \bibinfo{pages}{1273--1282}.
\newblock


\bibitem[Nguyen et~al\mbox{.}(2022)]%
        {nguyen2022federated}
\bibfield{author}{\bibinfo{person}{Dinh~C Nguyen}, \bibinfo{person}{Quoc-Viet
  Pham}, \bibinfo{person}{Pubudu~N Pathirana}, \bibinfo{person}{Ming Ding},
  \bibinfo{person}{Aruna Seneviratne}, \bibinfo{person}{Zihuai Lin},
  \bibinfo{person}{Octavia Dobre}, {and} \bibinfo{person}{Won-Joo Hwang}.}
  \bibinfo{year}{2022}\natexlab{}.
\newblock \showarticletitle{Federated learning for smart healthcare: A survey}.
\newblock \bibinfo{journal}{\emph{ACM Computing Surveys (CSUR)}}
  \bibinfo{volume}{55}, \bibinfo{number}{3} (\bibinfo{year}{2022}),
  \bibinfo{pages}{1--37}.
\newblock


\bibitem[Park and Tuzhilin(2008)]%
        {park2008long}
\bibfield{author}{\bibinfo{person}{Yoon-Joo Park} {and}
  \bibinfo{person}{Alexander Tuzhilin}.} \bibinfo{year}{2008}\natexlab{}.
\newblock \showarticletitle{The long tail of recommender systems and how to
  leverage it}. In \bibinfo{booktitle}{\emph{RecSys}}. \bibinfo{pages}{11--18}.
\newblock


\bibitem[Pennington et~al\mbox{.}(2014)]%
        {pennington2014glove}
\bibfield{author}{\bibinfo{person}{Jeffrey Pennington},
  \bibinfo{person}{Richard Socher}, {and} \bibinfo{person}{Christopher~D
  Manning}.} \bibinfo{year}{2014}\natexlab{}.
\newblock \showarticletitle{Glove: Global vectors for word representation}. In
  \bibinfo{booktitle}{\emph{EMNLP}}. \bibinfo{pages}{1532--1543}.
\newblock


\bibitem[Pillutla et~al\mbox{.}(2019)]%
        {pillutla2019robust}
\bibfield{author}{\bibinfo{person}{Krishna Pillutla}, \bibinfo{person}{Sham~M
  Kakade}, {and} \bibinfo{person}{Zaid Harchaoui}.}
  \bibinfo{year}{2019}\natexlab{}.
\newblock \showarticletitle{Robust aggregation for federated learning}.
\newblock \bibinfo{journal}{\emph{arXiv preprint arXiv:1912.13445}}
  (\bibinfo{year}{2019}).
\newblock


\bibitem[Portnoy et~al\mbox{.}(2020)]%
        {portnoy2020towards}
\bibfield{author}{\bibinfo{person}{Amit Portnoy}, \bibinfo{person}{Yoav
  Tirosh}, {and} \bibinfo{person}{Danny Hendler}.}
  \bibinfo{year}{2020}\natexlab{}.
\newblock \showarticletitle{Towards Federated Learning With Byzantine-Robust
  Client Weighting}.
\newblock \bibinfo{journal}{\emph{arXiv preprint arXiv:2004.04986}}
  (\bibinfo{year}{2020}).
\newblock


\bibitem[Ramaswamy et~al\mbox{.}(2019)]%
        {ramaswamy2019federated}
\bibfield{author}{\bibinfo{person}{Swaroop Ramaswamy}, \bibinfo{person}{Rajiv
  Mathews}, \bibinfo{person}{Kanishka Rao}, {and}
  \bibinfo{person}{Fran{\c{c}}oise Beaufays}.} \bibinfo{year}{2019}\natexlab{}.
\newblock \showarticletitle{Federated learning for emoji prediction in a mobile
  keyboard}.
\newblock \bibinfo{journal}{\emph{arXiv preprint arXiv:1906.04329}}
  (\bibinfo{year}{2019}).
\newblock


\bibitem[Reddi et~al\mbox{.}(2021)]%
        {reddi2021adaptive}
\bibfield{author}{\bibinfo{person}{Sashank~J. Reddi}, \bibinfo{person}{Zachary
  Charles}, \bibinfo{person}{Manzil Zaheer}, \bibinfo{person}{Zachary Garrett},
  \bibinfo{person}{Keith Rush}, \bibinfo{person}{Jakub Kone{\v{c}}n{\'y}},
  \bibinfo{person}{Sanjiv Kumar}, {and} \bibinfo{person}{Hugh~Brendan
  McMahan}.} \bibinfo{year}{2021}\natexlab{}.
\newblock \showarticletitle{Adaptive Federated Optimization}. In
  \bibinfo{booktitle}{\emph{ICLR}}.
\newblock


\bibitem[Reynolds et~al\mbox{.}(2009)]%
        {reynolds2009gaussian}
\bibfield{author}{\bibinfo{person}{Douglas~A Reynolds} {et~al\mbox{.}}}
  \bibinfo{year}{2009}\natexlab{}.
\newblock \showarticletitle{Gaussian mixture models.}
\newblock \bibinfo{journal}{\emph{Encyclopedia of biometrics}}
  \bibinfo{volume}{741}, \bibinfo{number}{659-663} (\bibinfo{year}{2009}).
\newblock


\bibitem[Rieke et~al\mbox{.}(2020)]%
        {rieke2020future}
\bibfield{author}{\bibinfo{person}{Nicola Rieke}, \bibinfo{person}{Jonny
  Hancox}, \bibinfo{person}{Wenqi Li}, \bibinfo{person}{Fausto Milletari},
  \bibinfo{person}{Holger~R Roth}, \bibinfo{person}{Shadi Albarqouni},
  \bibinfo{person}{Spyridon Bakas}, \bibinfo{person}{Mathieu~N Galtier},
  \bibinfo{person}{Bennett~A Landman}, \bibinfo{person}{Klaus Maier-Hein},
  {et~al\mbox{.}}} \bibinfo{year}{2020}\natexlab{}.
\newblock \showarticletitle{The future of digital health with federated
  learning}.
\newblock \bibinfo{journal}{\emph{NPJ digital medicine}} \bibinfo{volume}{3},
  \bibinfo{number}{1} (\bibinfo{year}{2020}), \bibinfo{pages}{1--7}.
\newblock


\bibitem[Ruder(2016)]%
        {ruder2016overview}
\bibfield{author}{\bibinfo{person}{Sebastian Ruder}.}
  \bibinfo{year}{2016}\natexlab{}.
\newblock \showarticletitle{An overview of gradient descent optimization
  algorithms}.
\newblock \bibinfo{journal}{\emph{arXiv preprint arXiv:1609.04747}}
  (\bibinfo{year}{2016}).
\newblock


\bibitem[Sheller et~al\mbox{.}(2020)]%
        {sheller2020federated}
\bibfield{author}{\bibinfo{person}{Micah~J Sheller}, \bibinfo{person}{Brandon
  Edwards}, \bibinfo{person}{G~Anthony Reina}, \bibinfo{person}{Jason Martin},
  \bibinfo{person}{Sarthak Pati}, \bibinfo{person}{Aikaterini Kotrotsou},
  \bibinfo{person}{Mikhail Milchenko}, \bibinfo{person}{Weilin Xu},
  \bibinfo{person}{Daniel Marcus}, \bibinfo{person}{Rivka~R Colen},
  {et~al\mbox{.}}} \bibinfo{year}{2020}\natexlab{}.
\newblock \showarticletitle{Federated learning in medicine: facilitating
  multi-institutional collaborations without sharing patient data}.
\newblock \bibinfo{journal}{\emph{Scientific reports}} \bibinfo{volume}{10},
  \bibinfo{number}{1} (\bibinfo{year}{2020}), \bibinfo{pages}{1--12}.
\newblock


\bibitem[Sun et~al\mbox{.}(2019)]%
        {sun2019can}
\bibfield{author}{\bibinfo{person}{Ziteng Sun}, \bibinfo{person}{Peter
  Kairouz}, \bibinfo{person}{Ananda~Theertha Suresh}, {and}
  \bibinfo{person}{H~Brendan McMahan}.} \bibinfo{year}{2019}\natexlab{}.
\newblock \showarticletitle{Can you really backdoor federated learning?}
\newblock \bibinfo{journal}{\emph{arXiv preprint arXiv:1911.07963}}
  (\bibinfo{year}{2019}).
\newblock


\bibitem[Wang et~al\mbox{.}(2020)]%
        {DBLP:conf/nips/WangSRVASLP20}
\bibfield{author}{\bibinfo{person}{Hongyi Wang}, \bibinfo{person}{Kartik
  Sreenivasan}, \bibinfo{person}{Shashank Rajput}, \bibinfo{person}{Harit
  Vishwakarma}, \bibinfo{person}{Saurabh Agarwal}, \bibinfo{person}{Jy yong
  Sohn}, \bibinfo{person}{Kangwook Lee}, {and} \bibinfo{person}{Dimitris~S.
  Papailiopoulos}.} \bibinfo{year}{2020}\natexlab{}.
\newblock \showarticletitle{Attack of the Tails: Yes, You Really Can Backdoor
  Federated Learning}. In \bibinfo{booktitle}{\emph{NIPS}}.
\newblock


\bibitem[Wu et~al\mbox{.}(2020)]%
        {wu-etal-2020-mind}
\bibfield{author}{\bibinfo{person}{Fangzhao Wu}, \bibinfo{person}{Ying Qiao},
  \bibinfo{person}{Jiun-Hung Chen}, \bibinfo{person}{Chuhan Wu},
  \bibinfo{person}{Tao Qi}, \bibinfo{person}{Jianxun Lian},
  \bibinfo{person}{Danyang Liu}, \bibinfo{person}{Xing Xie},
  \bibinfo{person}{Jianfeng Gao}, \bibinfo{person}{Winnie Wu}, {and}
  \bibinfo{person}{Ming Zhou}.} \bibinfo{year}{2020}\natexlab{}.
\newblock \showarticletitle{{MIND}: A Large-scale Dataset for News
  Recommendation}. In \bibinfo{booktitle}{\emph{ACL}}.
  \bibinfo{pages}{3597--3606}.
\newblock


\bibitem[Xie et~al\mbox{.}(2021)]%
        {xie2021crfl}
\bibfield{author}{\bibinfo{person}{Chulin Xie}, \bibinfo{person}{Minghao Chen},
  \bibinfo{person}{Pin-Yu Chen}, {and} \bibinfo{person}{Bo Li}.}
  \bibinfo{year}{2021}\natexlab{}.
\newblock \showarticletitle{Crfl: Certifiably robust federated learning against
  backdoor attacks}. In \bibinfo{booktitle}{\emph{ICML}}.
  \bibinfo{pages}{11372--11382}.
\newblock


\bibitem[Xie et~al\mbox{.}(2020)]%
        {Xie2020DBA:}
\bibfield{author}{\bibinfo{person}{Chulin Xie}, \bibinfo{person}{Keli Huang},
  \bibinfo{person}{Pin-Yu Chen}, {and} \bibinfo{person}{Bo Li}.}
  \bibinfo{year}{2020}\natexlab{}.
\newblock \showarticletitle{DBA: Distributed Backdoor Attacks against Federated
  Learning}. In \bibinfo{booktitle}{\emph{ICLR}}.
\newblock


\bibitem[Xu et~al\mbox{.}(2021)]%
        {xu2021federated}
\bibfield{author}{\bibinfo{person}{Jie Xu}, \bibinfo{person}{Benjamin~S
  Glicksberg}, \bibinfo{person}{Chang Su}, \bibinfo{person}{Peter Walker},
  \bibinfo{person}{Jiang Bian}, {and} \bibinfo{person}{Fei Wang}.}
  \bibinfo{year}{2021}\natexlab{}.
\newblock \showarticletitle{Federated learning for healthcare informatics}.
\newblock \bibinfo{journal}{\emph{Journal of Healthcare Informatics Research}}
  \bibinfo{volume}{5} (\bibinfo{year}{2021}), \bibinfo{pages}{1--19}.
\newblock


\bibitem[Yang et~al\mbox{.}(2018)]%
        {yang2018applied}
\bibfield{author}{\bibinfo{person}{Timothy Yang}, \bibinfo{person}{Galen
  Andrew}, \bibinfo{person}{Hubert Eichner}, \bibinfo{person}{Haicheng Sun},
  \bibinfo{person}{Wei Li}, \bibinfo{person}{Nicholas Kong},
  \bibinfo{person}{Daniel Ramage}, {and} \bibinfo{person}{Fran{\c{c}}oise
  Beaufays}.} \bibinfo{year}{2018}\natexlab{}.
\newblock \showarticletitle{Applied federated learning: Improving google
  keyboard query suggestions}.
\newblock \bibinfo{journal}{\emph{arXiv preprint arXiv:1812.02903}}
  (\bibinfo{year}{2018}).
\newblock


\bibitem[Yin et~al\mbox{.}(2018)]%
        {yin2018byzantine}
\bibfield{author}{\bibinfo{person}{Dong Yin}, \bibinfo{person}{Yudong Chen},
  \bibinfo{person}{Ramchandran Kannan}, {and} \bibinfo{person}{Peter
  Bartlett}.} \bibinfo{year}{2018}\natexlab{}.
\newblock \showarticletitle{Byzantine-robust distributed learning: Towards
  optimal statistical rates}. In \bibinfo{booktitle}{\emph{ICML}}.
  \bibinfo{pages}{5650--5659}.
\newblock


\bibitem[Zaheer et~al\mbox{.}(2018)]%
        {zaheer2018adaptive}
\bibfield{author}{\bibinfo{person}{Manzil Zaheer}, \bibinfo{person}{Sashank
  Reddi}, \bibinfo{person}{Devendra Sachan}, \bibinfo{person}{Satyen Kale},
  {and} \bibinfo{person}{Sanjiv Kumar}.} \bibinfo{year}{2018}\natexlab{}.
\newblock \showarticletitle{Adaptive methods for nonconvex optimization}.
\newblock \bibinfo{journal}{\emph{NIPS}}  \bibinfo{volume}{31}
  (\bibinfo{year}{2018}).
\newblock


\bibitem[Zhang et~al\mbox{.}(2017)]%
        {Zhang_2017_ICCV}
\bibfield{author}{\bibinfo{person}{Xiao Zhang}, \bibinfo{person}{Zhiyuan Fang},
  \bibinfo{person}{Yandong Wen}, \bibinfo{person}{Zhifeng Li}, {and}
  \bibinfo{person}{Yu Qiao}.} \bibinfo{year}{2017}\natexlab{}.
\newblock \showarticletitle{Range Loss for Deep Face Recognition With
  Long-Tailed Training Data}. In \bibinfo{booktitle}{\emph{ICCV}}.
  \bibinfo{pages}{5419--5428}.
\newblock


\bibitem[Zhong et~al\mbox{.}(2019)]%
        {Zhong_2019_CVPR}
\bibfield{author}{\bibinfo{person}{Yaoyao Zhong}, \bibinfo{person}{Weihong
  Deng}, \bibinfo{person}{Mei Wang}, \bibinfo{person}{Jiani Hu},
  \bibinfo{person}{Jianteng Peng}, \bibinfo{person}{Xunqiang Tao}, {and}
  \bibinfo{person}{Yaohai Huang}.} \bibinfo{year}{2019}\natexlab{}.
\newblock \showarticletitle{Unequal-Training for Deep Face Recognition With
  Long-Tailed Noisy Data}. In \bibinfo{booktitle}{\emph{CVPR}}.
  \bibinfo{pages}{7804--7813}.
\newblock


\end{thebibliography}
